\documentclass[preprint,aps,amsmath,amssymb,nofootinbib,12pt]{revtex4}

\usepackage{epsfig}
\usepackage{slashed}
\usepackage{graphicx}
\usepackage{multirow,color}
\usepackage{amsmath}
\usepackage{float}
\usepackage{diagbox}
\usepackage{CJK}
\usepackage{color}
\usepackage{xcolor}
\usepackage{times}
\usepackage{subfigure}
\usepackage{bm}
\usepackage{braket}
\usepackage{booktabs}
\usepackage{array}
\usepackage[mathscr]{euscript}
\usepackage{caption}
\usepackage{makecell}
\usepackage{epstopdf}
\usepackage{hyperref}

\makeatletter

\newcommand{\Rmnum}[1]{\expandafter\@slowromancap\romannumeral #1@}
\makeatother
\graphicspath{{fig/}}
\begin{document}
\title{Probing the couplings of an axion-like particle with leptons via three-lepton final state processes at future $\boldsymbol{e^{-}p}$ colliders}
\author{Chong-Xing Yue$^{1,2}$}
\thanks{cxyue@lnnu.edu.cn}
\author{Xin-Yang Li$^{1,2}$}
\thanks{lxy91108@163.com~(Corresponding author)}
\author{Mei-Shu-Yu Wang$^{1,2}$}
\thanks{13904039103@163.com}
\author{Yang-Yang Bu$^{1,2}$}
\thanks{byy20011020@163.com}

\affiliation{
$^1$Department of Physics, Liaoning Normal University, Dalian 116029, China\\
$^2$Center for Theoretical and Experimental High Energy Physics, Liaoning Normal University, China
}

\begin{abstract}
 The axion-like particle~(ALP) is one of the best motivated particles beyond the Standard Model~(SM). We explore the possibility of detecting the couplings of ALP with leptons via three-lepton final state processes $e^- p \to e^- j a~(a \to \ell^+ \ell^-)$ at the LHeC (FCC-eh). For completeness, we investigate the cases where the ALP decays not only into electron and muon pairs but also into tau pairs, and perform a detailed simulation for the decays of taus. We find that the prospective sensitivities of the LHeC (FCC-eh) with the center-of-mass energy $\sqrt{s}=1.3~(3.5)$ TeV and the integrated luminosity $\mathcal{L}=$ $1~(2)$ ab$^{-1}$ to the ALP-lepton couplings are not only stronger than some existing bounds of the LHC and LEP but also complementary to the expected bounds of the CEPC and FCC-ee.
\end{abstract}
\maketitle

\section{Introduction}
The Standard Model (SM) of particle physics has enjoyed remarkable success in describing the fundamental constituents of matter and their interactions. However, there are many reasons to believe that the SM is not an ultimate theory of nature, with numerous phenomena such as dark matter~\cite{Duffy:2009ig,Chadha-Day:2021szb}, gauge hierarchy problem~\cite{Feng:2013pwa}, the matter-antimatter asymmetry~\cite{DiBari:2021fhs} as well as the origin of neutrino mass~\cite{Gonzalez-Garcia:2007dlo} remaining unexplained. These unresolved questions suggest that there must be new physics beyond the SM. Axion-like particles~(ALPs, denoted as ``$a$") appear in many extensions of the SM, potentially playing a key role in addressing some of these outstanding puzzles. ALPs are $CP$-odd pseudo Nambu-Goldstone bosons~(PNGBs) associated with spontaneously broken global symmetries at high scales, generalizing the idea of the QCD axions proposed to solve the strong $CP$ problem~\cite{Peccei:1977hh,Peccei:1977ur,Weinberg:1977ma,Dine:1981rt} via Peccei-Quinn~(PQ) mechanism. Unlike the QCD axions, ALPs are independent of PQ mechanism and enjoy less model constraints than the QCD axions, allowing their masses and couplings to vary independently. This flexibility expands their parameter space, resulting in a wide range of rich phenomena in low- and high-energy experiments.

The masses and the couplings of ALPs can extend over many orders of magnitude, which have been widely studied using various experimental datas from particle physics, astrophysics and cosmology. Astrophysical and cosmological observations can put strong constraints on light ALPs with masses in the range from a few eV to MeV, such as big bang nucleosynthesis~(BBN)~\cite{Ghosh:2020vti,Depta:2020zbh}, cosmic microwave background~(CMB)~\cite{Marsh:2015xka,Ejlli:2013uda,Planck:2015fie} and supernova 1987a~\cite{Caputo:2024oqc,Ferreira:2022xlw,Diamond:2023scc,Caputo:2022mah}. As the masses of ALPs increase beyond the MeV scale, the focus of their study shifts to more direct experimental probes within the realm of particle physics. Numerous studies in the intensity frontier have
explored the couplings of ALPs with masses in the range from a few MeV to several GeV, such as beam dump and low-energy $e^+e^-$collider experiments~\cite{NA64:2020qwq,Riordan:1987aw,Dolan:2017osp,CHARM:1985anb,Blumlein:1990ay,
Waites:2022tov,Belle-II:2020jti,Alda:2024cxn}.
For heavier ALPs, with masses ranging from a few GeV to the TeV scale, high-energy colliders become indispensable. A variety of dedicated high-energy collider experiments, such as the LEP, LHC and future facilities, have been proposed to probe the couplings of these heavier ALPs with gluons, quarks and gauge bosons~\cite{Phan:2023dqw,Anuar:2024myn,Ghebretinsaea:2022djg,Mimasu:2014nea,Cheung:2024qge,Haghighat:2020nuh,
Bauer:2018uxu,Zhang:2021sio,Wang:2022ock,Yue:2022ash,Yue:2021iiu,Cheung:2023nzg,Balkin:2023gya,Gao:2024rgl}.

Up to now, most phenomenological analyses of ALPs have concentrated on their couplings to gluons, quarks and gauge bosons. However, ALPs could also interact with the SM leptons and cause novel signatures at collider experiments. Several interesting studies on the ALP-lepton couplings already exist. For example, Refs.~\cite{Jiang:2024cqj,Lu:2022zbe,Lu:2023ryd} have explored the four-point interactions~$(W\text{-}\ell\text{-}\nu\text{-}a)$~\cite{Altmannshofer:2022ckw} for leptophilic ALPs at high-energy colliders. In addition, many studies have focused on the three-point interactions between ALP and the SM leptons, such as those explored in Refs.\cite{Calibbi:2022izs, Bauer:2018uxu}, which examine direct probes of the ALP-lepton couplings via the process $p p \to h \to Z a~(a a) \to 4 \ell$ and $e^+ e^- \to Z \to \ell^+ \ell^- a$ at the future $p p$ and $e^+e^-$ colliders. The couplings of ALP with lepton pairs, $a\ell\bar{\ell}$, are proportional to the lepton mass and generally highly suppressed, but they can induce the couplings between the ALP and electroweak~(EW) gauge bosons at one-loop. Consequently, these loop-induced ALP-EW gauge boson couplings generate rich phenomena at high-energy colliders, which can be translated into novel signatures of the $a\ell\bar{\ell}$ couplings in high-energy collider experiments. References~\cite{Bonilla:2023dtf,Yue:2024xrc} have explored the possibility of detecting the $a\ell\bar{\ell}$ couplings from their loop-level impact on the ALP couplings to EW gauge bosons $\gamma\gamma$, $W^+ W^-$, $ZZ$ and $Z\gamma$ at the $p p$ and $e^+ e^-$ colliders.

The $e^-p$ colliders, such as the Large Hadron electron Collider~(LHeC)~\cite{Klein:2009qt,LHeCStudyGroup:2012zhm,Bruening:2013bga,LHeC:2020van} and Future Circular Collider-electron hadron ~(FCC-eh)~\cite{FCC:2018byv,Zimmermann:2014qxa}, have been proposed to combine the excellent performance of the $p p$ and $e^+ e^-$  colliders. By complementing the proton ring with an electron beam, they enable deep inelastic scattering~(DIS) of electrons and protons at TeV energies. Compared to the $e^+ e^-$ colliders, the $e^-p$ colliders achieve higher center-of-mass energies, allowing for the production of heavier ALPs and the exploration of these heavier ALPs. At the same time, the $e^-p$ collider can offer a clean environment
with suppressed backgrounds from strong interaction, enabling high precision measurements compared to the $p p$ colliders. Additionally, the asymmetric initial state allows for the separation of backward and forward scattering, which can greatly increase the significance of the signal and get opportunities that cannot be observed at the $p p$ colliders. In summary, the $e^- p$ colliders serve as an excellent complement to both the $e^+ e^-$ and $pp$ colliders. Therefore, $e^- p$ colliders exhibit significant potential for probing the ALP-lepton couplings. Considering that previous studies on the ALP-lepton couplings have primarily focused on the $p p$ and $e^+ e^-$ colliders, we now turn to exploring the possibility of detecting the ALP-lepton couplings at the $e^-p$ colliders.

In this work, we implement the investigation of the ALP-lepton couplings through three-lepton final state processes $e^- p \to e^- j a~(a \to \ell^+ \ell^-$ with $\ell$ being $\mu$ or $\tau)$ at the LHeC~(FCC-eh). The LHeC (FCC-eh) is expected to operate at the center-of-mass energy $\sqrt{s}=1.3~(3.5)$ TeV with the integrated luminosity $\mathcal{L}=$ $1$~(2) ab$^{-1}$, and corresponding electron and proton beam energies of 60 GeV and 7~(50) TeV, respectively. Comparing our prospective sensitivities to the ALP-lepton couplings with existing experimental bounds and projected experimental limits from other studies, we find that the prospective sensitivities of the LHeC (FCC-eh) are not covered by the exclusion regions given by current and future experiments within a specific mass range.

The structure of the paper is as follows. In Sec. II, we present a summary of the effective Lagrangian describing the interactions of ALP with the SM particles and present the branching ratios for different decay modes of ALP. Section III provides a detailed analysis for the possibility of detecting the couplings of ALP with leptons through three-lepton final state processes $e^- p \to e^- j a~(a \to \mu^+ \mu^-$) and $e^{-}p\rightarrow e^-ja~(a \to \tau^+ \tau^-, \tau^+ \tau^- \to \ell^+ \ell^{-} \slashed E, \ell= e,\mu)$ based on the LHeC and FCC-eh detector simulation and compares our results with existing experimental limits as well as the projected limits from future experiments. Finally, the conclusions are summarized in Sec. IV.

\section{The theory framework}

The interactions of ALP with the SM particles can be described by the effective Lagrangian that includes operators with dimension up to five~\cite{Georgi:1986df,Bonilla:2021ufe,Bonilla:2022qgm,Choi:1986zw}

\begin{eqnarray}
\begin{split}
\label{eq:2.1}
\mathcal{L}_{\text{ALP}}
	= \frac{1}{4} \partial_\mu a \partial^\mu a - \frac{1}{2} m_a^2 a^2 +  \mathcal{L}_a^{gauge} + \mathcal{L}_{\partial a}^{fermion}.
\end{split}
\end{eqnarray}
At energies above the electroweak symmetry breaking scale, the ALP-gauge boson effective couplings, represented by $\mathcal{L}_a^{gauge}$, can be written as
\begin{eqnarray}
\begin{split}
\label{eq:2.2}
\mathcal{L}_a^{gauge}
	= - C_{\tilde{G}} \frac{a}{f_a} G^{\alpha}_{\mu\nu} \tilde{G}^{\mu\nu,\alpha} - C_{\tilde{W}} \frac{a}{f_a} W^{i}_{\mu\nu} \tilde{W}^{\mu\nu,i} - C_{\tilde{B}} \frac{a}{f_a} B_{\mu\nu} \tilde{B}^{\mu\nu}.
\end{split}
\end{eqnarray}
Here $G^{\alpha}_{\mu\nu}$, $W^{i}_{\mu\nu}$ and $B_{\mu\nu}$ represent the field strength tensors for the gauge groups $SU(3)_{C}$, $SU(2)_{L}$ and $U(1)_{Y}$, respectively. $\tilde{G}^{\mu\nu,\alpha}$, $\tilde{W}^{\mu\nu,i}$ and $\tilde{B}^{\mu\nu}$ are the corresponding dual field strength tensors, which are defined as $\tilde{V}^{\mu\nu}=\frac{1}{2}\epsilon^{\mu\nu\lambda\kappa}V_{\lambda\kappa}$ $(V \in G$, $W$, $B)$ with $\epsilon^{0123}= 1$. $C_{\tilde{G}}$, $C_{\tilde{W}}$ and $ C_{\tilde{B}}$ denote the respective coupling constants.
$\mathcal{L}_{\partial a}^{fermion}$ represents the gauge-invariant ALP-fermion derivative interactions,
\begin{eqnarray}
\begin{split}
\label{eq:2.3}
\mathcal{L}_{\partial a}^{fermion}
	= \sum_{\substack{\psi}} \frac{\partial_{\mu} a}{f_a}  \bar\psi \gamma^\mu \mathbf{c}_\psi \psi,
\end{split}
\end{eqnarray}
where $\mathbf{c}_\psi$ are Hermitian matrices in flavor space. In this paper, we focus on purely leptonic couplings, then the gauge-invariant ALP-fermion derivative interactions can be written as~\cite{Bonilla:2023dtf}

\begin{equation} \label{eq:2.4}
\begin{aligned}
      \mathcal{L}&_{\partial a}^{fermion}  = \frac{\partial_\mu a}{f_a} \, \bar{L}_L \gamma^\mu c_L L_L + \frac{\partial_\mu a}{f_a} \, \bar{\ell}_R \gamma^\mu c_R \ell_R \,, \\
        & = \frac{\partial_\mu a}{2 f_a} \, \bar{\nu}_L \gamma^\mu (1 - \gamma^5) c_L \nu_{\ell} + \frac{\partial_\mu a}{2 f_a} \, \bar{\ell} \gamma^\mu (( c_R + c_L ) + ( c_R - c_L ) \gamma^5 ) \ell \,.
\end{aligned}
\end{equation}
Here $P_{R,L}$ stand for the chirality projectors, the left-handed EW lepton fields are decomposed into their charged and neutral components $L_L=(\ell_L, \nu_L)$, with $\ell= \ell_R + \ell_L$.

In the flavor-diagonal scenario, the ``phenomenological" couplings to the SM leptons are determined solely by the axial component of the operators in Eq.~\eqref{eq:2.4}~\cite{Bonilla:2021ufe,Bonilla:2022qgm}. As a result, the ALP-lepton effective interactions can be expressed as \footnote[1]{In Eq.~\eqref{eq:2.5}, the vectorial coupling vanishes due to the conservation of vectorial currents at classical level. However, this does not hold for flavor nondiagonal couplings, which will not be considered in this work.}
\begin{equation}\label{eq:2.5}
\begin{aligned}
    \mathcal{L}_{\partial a}^{fermion}  \supset  \frac{\partial_\mu a}{f_a} \sum_{\ell} (g_{a\nu\nu})_{\ell\ell} \overline{\nu}_\ell \gamma^\mu \gamma^5 \nu_\ell
     + \frac{\partial_\mu a}{f_a} \sum_{\ell} (g_{a\ell\ell})_{\ell\ell} \overline{\ell} \gamma^\mu \gamma^5 \ell \,,
 \end{aligned}
\end{equation}
with
\begin{equation} \label{eq:2.6}
    g_{a\nu\nu} \equiv (- c_L) \,, \qquad g_{a\ell\ell} \equiv (c_R - c_L) \,.
\end{equation}
where $g_{a\nu\nu}$ and $g_{a\ell\ell}$ represent the coupling coefficients of ALP to neutrinos and charged leptons, respectively.

After electroweak symmetry breaking, the ALP-gauge boson effective couplings presented in Eq.~(\ref{eq:2.2}) can be rewritten in terms of the mass eigenstates of EW gauge bosons,
\begin{equation} \label{eq:2.7}
\begin{aligned}
 \mathcal{L}_a^{gauge}\,= & -\frac{1}{4} g_{agg} \,a\, G_{\mu \nu}^\alpha \widetilde{G}^{\mu \nu,\, \alpha}  - \frac{1}{4} g_{a\gamma\gamma} \,a\,F_{\mu \nu} \widetilde{F}^{\mu \nu} \\
    & - \frac{1}{4} g_{a\gamma Z} \,a\,F_{\mu \nu} \widetilde{Z}^{\mu \nu} - \frac{1}{4} g_{aZZ} \,a\,Z_{\mu \nu} \widetilde{Z}^{\mu \nu} - \frac{1}{2} g_{aWW} \,a\,W_{\mu \nu}^+ \widetilde{W}^{\mu \nu,\,-} \,,
\end{aligned}
\end{equation}
with
\begin{equation} \label{eq:2.8}
    \begin{aligned}
        & g_{agg} \equiv \frac{4}{f_a} c_{\tilde{G}} \,,  \qquad g_{a\gamma\gamma}  \equiv \frac{4}{f_a} \left( c_w^2 c_{\tilde{B}} + s_w^2 c_{\tilde{W}} \right) \,,   & \\
        & g_{aWW}   \equiv \frac{4}{f_a} c_{\tilde{W}} \,,  \qquad  g_{aZZ}  \equiv \frac{4}{f_a} \left( s_w^2 c_{\tilde{B}} + c_w^2 c_{\tilde{W}} \right) \,,  \qquad  g_{a\gamma Z} \equiv \frac{8}{f_a} c_w s_w \left( c_{\tilde{W}} - c_{\tilde{B}} \right) \,.
    \end{aligned}
\end{equation}
Here, $c_w (s_w)$ is the cosine (sine) of the weak mixing angle $\theta_{w}$.

\begin{figure}[!ht]
\centerline{
\includegraphics[width=0.4\textwidth]{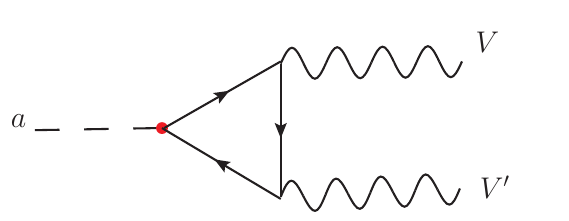}}
\vspace*{8pt}
\caption{One-loop diagrams that contribute to the effective ALP-EW gauge boson couplings, arising from the ALP-lepton couplings.
\protect\label{fig:1}}
\end{figure}

Based on the preceding discussion, it is evident that the ALP-lepton couplings give rise to the ALP-EW gauge boson effective couplings at one-loop, as extensively studied in Ref.~\cite{Bonilla:2023dtf}. The relevant Feynman diagrams are presented in Fig.~\ref{fig:1}, where $V$ and $V'$ denote the EW gauge bosons $\gamma$, $Z$ or $W$, meanwhile the leptons circulating in the loop may either be charged leptons or neutrinos, depending on the specific gauge bosons $V$ and $V'$. The contributions can be categorized into two distinct parts, the first is the chiral anomalous contribution, which is independent of lepton mass, and the second is a mass-dependent term. Since the masses of the $Z$ and $W$ bosons are much larger than those of the SM leptons, the mass-dependent terms contribute at most corrections of order $\mathcal{O} (m_{\ell}^2 / M_Z^2)$ for $g_{a\gamma Z}^{eff}$ and $g_{aZZ}^{eff}$, which are sufficiently small to be safely ignored. Furthermore, the effective ALP-photon coupling $g_{a\gamma \gamma}^{eff}$ depends solely on the ratio $m_{\ell}^2/m_a^2$. When $m_a \gg m_{\ell}$, the mass-dependent terms become negligible. For a more detailed discussion, see Ref.~\cite{Bonilla:2023dtf}. When the ALP mass $m_a$ exceeds the mass $m_{\ell}$ of the lepton running in the loop, the ALP-EW gauge boson effective couplings induced by the ALP-lepton couplings can be approximated as

\begin{equation}\label{eq:2.9}
\begin{aligned}
 &
 g_{a\gamma\gamma}^{eff} \approx - \frac{\alpha_{em}}{\pi f_a}Tr(g_{a\ell\ell}),
 \\
 &
 g_{a\gamma Z}^{eff} \approx - \frac{\alpha_{em}}{s_w c_w \pi f_a}[2 s_w^2 Tr(g_{a\ell\ell})  - Tr(g_{a\nu\nu})],
 \\
 &
 g_{aZZ}^{eff} \approx - \frac{\alpha_{em}}{2 s_w^2 c_w^2 \pi f_a}[2 s_w^4 Tr(g_{a\ell\ell})+ (1- 2 s_w^2)Tr(g_{a\nu\nu})],
\end{aligned}
\end{equation}
with
\begin{equation} \label{eq:2.10}
    \begin{aligned}
       \alpha_{em}=\frac{e^2}{4\pi} = \frac{g^2 g'^2}{4\pi(g^2+g'^2)} \,.
    \end{aligned}
\end{equation}

 From Eq~.\eqref{eq:2.9}, it can be observed that the experimental constraints on the ALP-photon effective coupling $g_{a\gamma \gamma}^{eff}$ can impose limits on the coupling $Tr(g_{a\ell\ell})$ alone, while the experimental bounds on the ALP-$\gamma Z$ and ALP-$ZZ$ effective couplings $g_{a\gamma Z}^{eff}$ and $g_{aZZ}^{eff}$ can be directly translated into constraints on two distinct combinations of the couplings defined in Eq.~\eqref{eq:2.6}.

According to the above analysis, it can be known that the ALP can decay into pairs of charged leptons and EW gauge bosons. When the ALP mass is below $100$ GeV, the decay channels $a \to \gamma\gamma$, $\ell^+ \ell^-$ and $Z \gamma$ become kinematically accessible. The corresponding decay widths are given by

\begin{equation}\label{eq:2.11}
\begin{aligned}
 &
 \Gamma(a\to\gamma\gamma) =  \frac{m_a^3}{64\pi}|g_{a\gamma\gamma}^{eff}|^2,
 \\
 &
  \Gamma(a\to\ell^{+}\ell^{-}) =  \frac{m_a m_{\ell}^2|g_{a\ell\ell}|^2}{8\pi f_a^2}\sqrt{1-\frac{4m_{\ell}^2}{m_a^2}},
 \\
 &
  \Gamma(a\to Z \gamma) =  \frac{m_a^3}{128\pi}|g_{a Z \gamma}^{eff}|^2 (1-\frac{m_Z^2}{m_a^2})^3.
\end{aligned}
\end{equation}

\begin{figure}[H]
\begin{center}
\centering\includegraphics [scale=0.3] {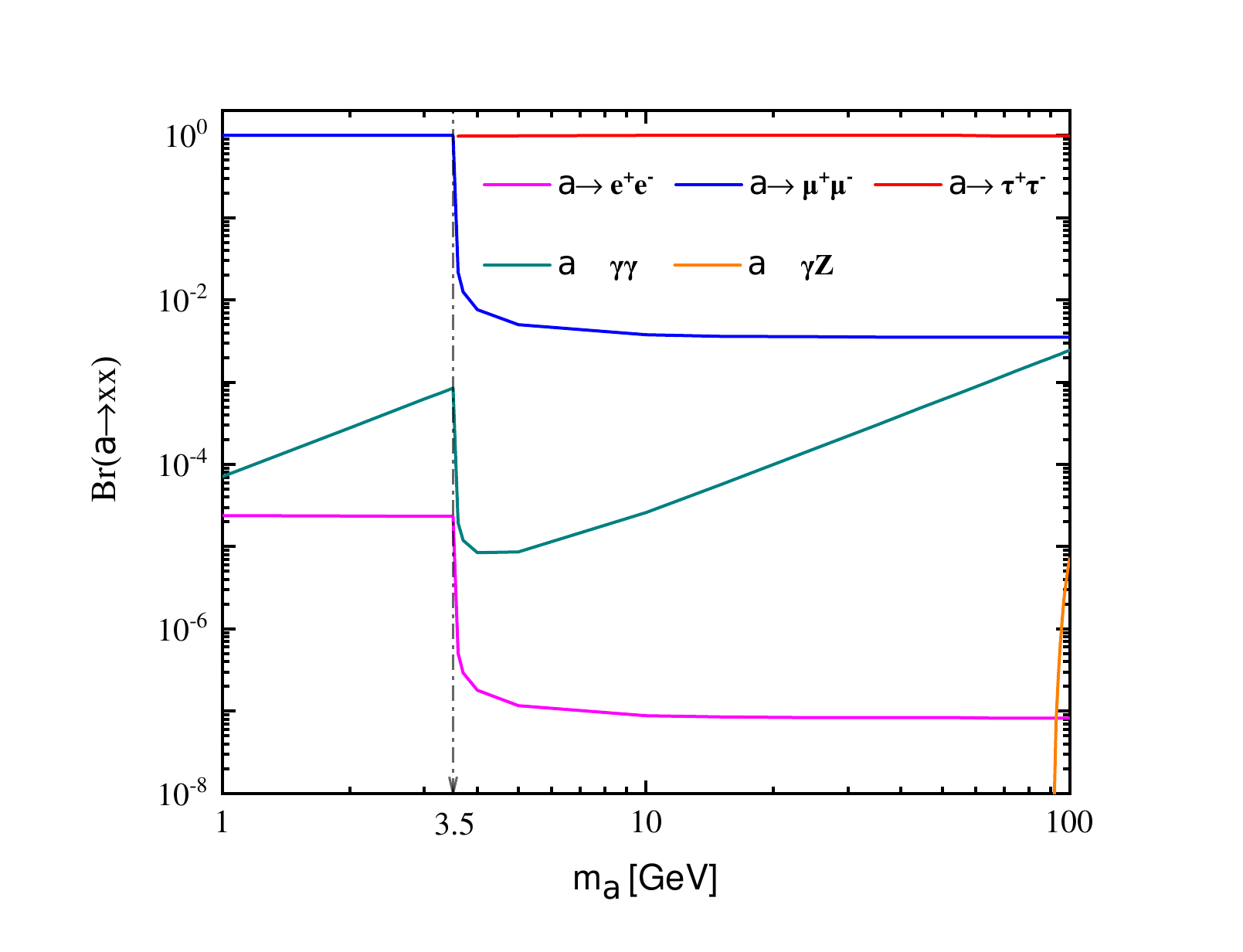}
\caption{The branching ratios for different ALP decay modes as functions of its mass $m_a$}
\label{fig:2}
\end{center}
\end{figure}

From above equations, one can see that, for a very light ALP with the mass below the electron mass threshold, the only allowed decay channel is $a\to \gamma\gamma$. As the ALP mass increases and exceeds the tau mass threshold~(approximately $3.5$ GeV), the decay channels involving charged leptons become accessible. However, the decay channel~$a \to \tau^+\tau^-$ takes over as the dominant channel, while the decays to lighter lepton pairs, $a\to e^+ e^-$ and $\mu^+\mu^-$, are suppressed due to their small masses. When the ALP mass $m_a$ exceeds $91$ GeV, the $Z\gamma$ mode becomes accessible, nevertheless, its branching ratio remains very small.
The specific branching ratios for these decay modes are shown in Fig.~\ref{fig:2}.

Consequently, we focus on probing the ALP-lepton couplings through the process $e^- p \to e^- j a $~($a\to \mu^+ \mu^-$) for the ALPs with masses ranging from $1$ to $5$ GeV. For the ALP masses above $5$ GeV, we shift our attention to the $\tau^+ \tau^-$ final state to investigate the ALP-lepton couplings due to its large branching ratio. Additionally, we seek to explore previously unexplored regions for the couplings of heavier ALPs to the SM leptons. To this end, we focus on the study of the $\tau^+ \tau^- $ final state for the ALPs with masses in the range of $5$ $\sim$ $500$ GeV. Since tau leptons are unstable, they decay either leptonically into electrons or muons with missing energy or hadronically into mesons with missing energy. Hadronic decays are often contaminated by significant background from other SM processes, such as the QCD interactions and hadronization effects, making it more challenging to distinguish the signal from the SM background. In contrast, leptonic decays of taus result in cleaner final states with less background interference. Therefore, we focus on the leptonic decays of taus in this study, as they offer clearer signals and allow for more precise analysis. Thus, we focus on probing the ALP-lepton couplings through three-lepton final state processes $e^- p \to e^- j a, a \to \mu^+ \mu^-$~($a \to \tau^+ \tau^-, \tau^+ \tau^- \to \ell^+ \ell^{-} \slashed E, \ell= e,\mu$) for the ALPs with masses ranging from $1(5)$ to $5~(500)$ GeV at the LHeC~(FCC-eh) with $\sqrt{s}=1.3~(3.5)$ TeV and $\mathcal{L}=$ $1$~(2) ab$^{-1}$. In this context, the ALP mass is much larger than the SM lepton mass $m_{\ell}$ for the corresponding signal process. So the signal can be analyzed using Eq.~\eqref{eq:2.9}.

\begin{figure}[H]
\begin{center}
\subfigure[]{\includegraphics [scale=0.35] {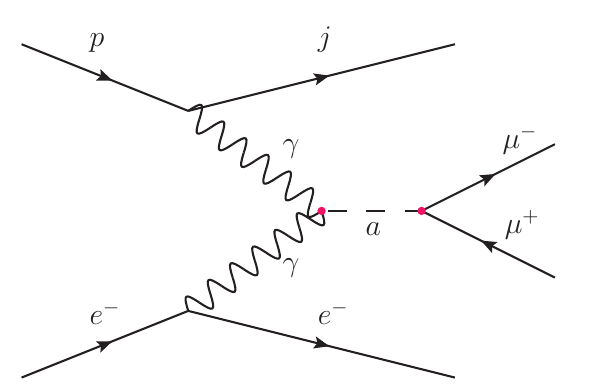}}
\subfigure[]{\includegraphics [scale=0.35] {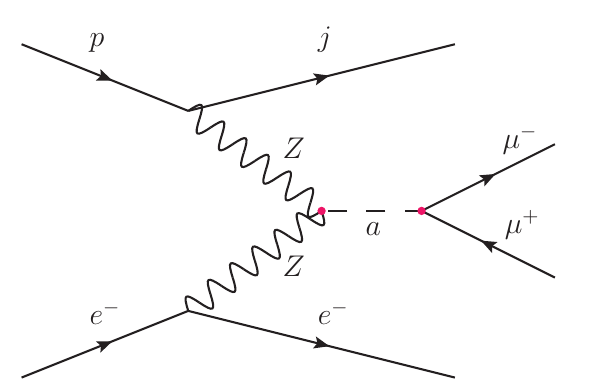}}
\subfigure[]{\includegraphics [scale=0.35] {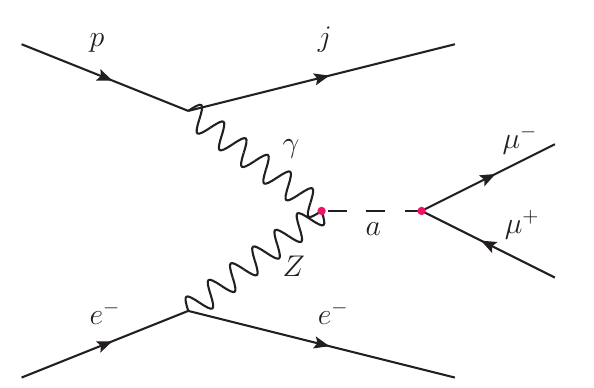}}
\caption{The Feynman diagrams for three-lepton final state process $e^- p \to e^- j a~(a \to \mu^+ \mu^-)$.}
\label{fig:3}
\end{center}
\end{figure}

\begin{figure}[H]
\begin{center}
\subfigure[]{\includegraphics [scale=0.35] {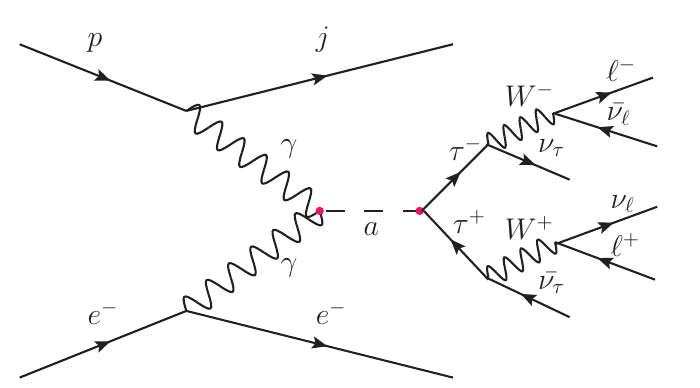}}
\subfigure[]{\includegraphics [scale=0.35] {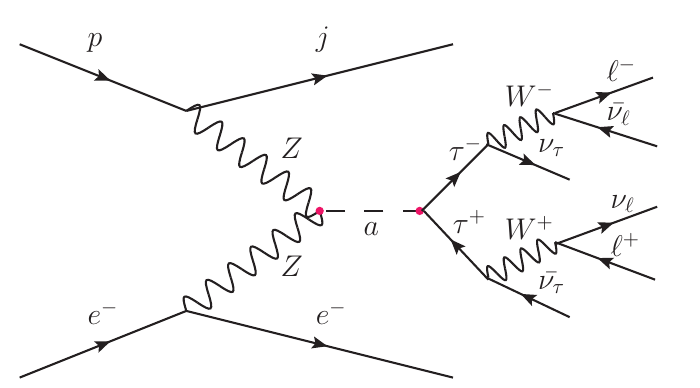}}
\subfigure[]{\includegraphics [scale=0.35] {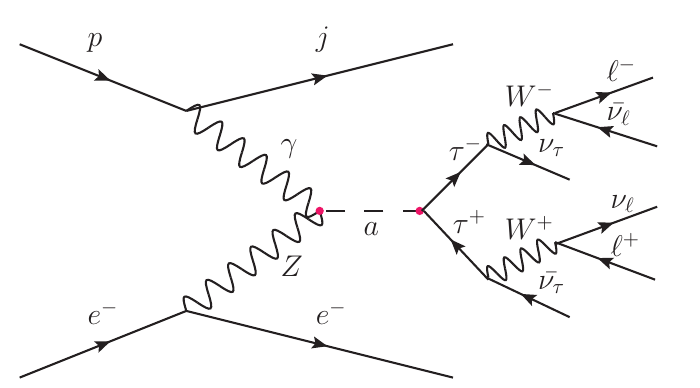}}
\caption{The Feynman diagrams for three-lepton final state process $e^{-}p\rightarrow e^-ja~(a \to \tau^+ \tau^-, \tau^+ \tau^- \to \ell^+ \ell^{-} \slashed E, \ell= e,\mu)$.}
\label{fig:4}
\end{center}
\end{figure}

The Feynman diagrams for the processes $e^- p \to e^- j a~(a \to \mu^+ \mu^-$) and $e^{-}p\rightarrow e^-ja~(a \to \tau^+ \tau^-, \tau^+ \tau^- \to \ell^+ \ell^{-} \slashed E, \ell= e,\mu)$ are displayed in Figs.~\ref{fig:3} and~\ref{fig:4}, respectively. For $c_{R}=0$, ALP has the same coupling strength to charged leptons and neutrinos, i.e, $Tr(g_{a\ell\ell})=Tr(g_{a\nu\nu})=-c_L$, in which case the ALP-lepton couplings can be explored via all the channels in these figures. For $c_{L}=0$, there is no ALP-neutrino couplings, i.e, $Tr(g_{a\nu\nu})=0$. In this case, the ALP-lepton couplings can still be explored through the same channels as the $c_{R}=0$ case, but with different expressions for the effective coupling coefficients $g_{a\gamma\gamma}^{eff}$ and $g_{a\gamma Z}^{eff}$, as shown in Eq.~\eqref{eq:2.9}.
For $c_{L}=c_{R}$, there is no tree-level flavor-diagonal ALP-charged lepton interactions, i.e, $Tr(g_{a\ell\ell})=0$, meaning the ALP cannot couple to two photons, which is known as photophobic ALP~\cite{Craig:2018kne}. However, this is not relevant to the study of three-lepton final state processes discussed in this paper, and therefore this scenario will be excluded from further consideration. Now, we will explore the possibility of detecting the couplings of ALP with leptons through three-lepton final state processes for the two cases, i.e. $c_{R}=0$ and $c_{L}=0$.

\section{The possibility of detecting the couplings of ALP with leptons at the LHeC and FCC-eh}
The model file for the effective Lagrangian is generated using \verb"FeynRules"~\cite{Alloul:2013bka}. A Monte Carlo~(MC) simulation is then carried out to assess the feasibility of detecting the ALP-lepton couplings at the LHeC and FCC-eh. It is well known that the polarization of the primary state electrons in the $e^- p$ colliders can affect the size of the production cross section. After comparison we find that the beam polarization as $P(e^{-}) = -80$ $\%$ can maximize the production cross section. So, in the rest of the calculations, the electron beam polarization is taken as $P(e^{-}) = -80$ $\%$. All signal and background events involving three leptons and a light jet ($u$, $d$, $c$, or $s$) to be discussed at the LHeC and FCC-eh in the next subsections will be generated using \verb"MadGraph5_aMC@NLO"~\cite{Alwall:2014hca} with basic cuts, which require leptons (electron and muon) with transverse momentum $p_T^l > 10$ GeV and the jet ($u$, $d$, $c$, or $s$) with transverse momentum $p_{T}^{j}>20$ GeV. The absolute values of the leptons pseudorapidity $\eta_{l}$ and the jet pseudorapidity $\eta_{j}$ need to be less than $2.5$ and $5$, respectively. Additionally, the angular separation requirements are $\Delta R _{ll}>0.4$ as well as $\Delta R _{lj}>0.4$ for leptons and jets, with $\Delta R$ defined as $\sqrt{(\Delta \phi)^{2}+(\Delta \eta)^{2}}$, where $\Delta \phi$ and $\Delta \eta$ are the azimuth difference and pseudorapidity difference of the lepton
pair or between a jet and a lepton, respectively. For all event generations, we use the \texttt{NN23LO1} parton distribution functions (PDFs)~\cite{Ball:2012cx,Deans:2013mha}. Both the factorization and renormalization scales for the signal and background simulations are set to the default dynamic scales provided by \texttt{MadGraph5}.
The \verb"PYTHIA8"~\cite{Sjostrand:2014zea} program is used for parton showering and hadronization and the \verb"DELPHES"~\cite{deFavereau:2013fsa} provides a fast simulation of the LHeC and FCC-eh detectors. Finally, kinematic and cut-based analyses are performed using \verb"MadAnalysis5"~\cite{Conte:2012fm,Conte:2014zja,Conte:2018vmg}.

\subsection{Searching for the ALP-lepton couplings via three-lepton final state processes for $c_{R}=0$}

\subsubsection{ The process $e^{-}p\rightarrow e^-ja~(a\rightarrow\mu^{+}\mu^{-})$}\label{subsec:first}

\begin{figure}[H]
\begin{center}
\subfigure[]{\includegraphics [width=0.52\textwidth] {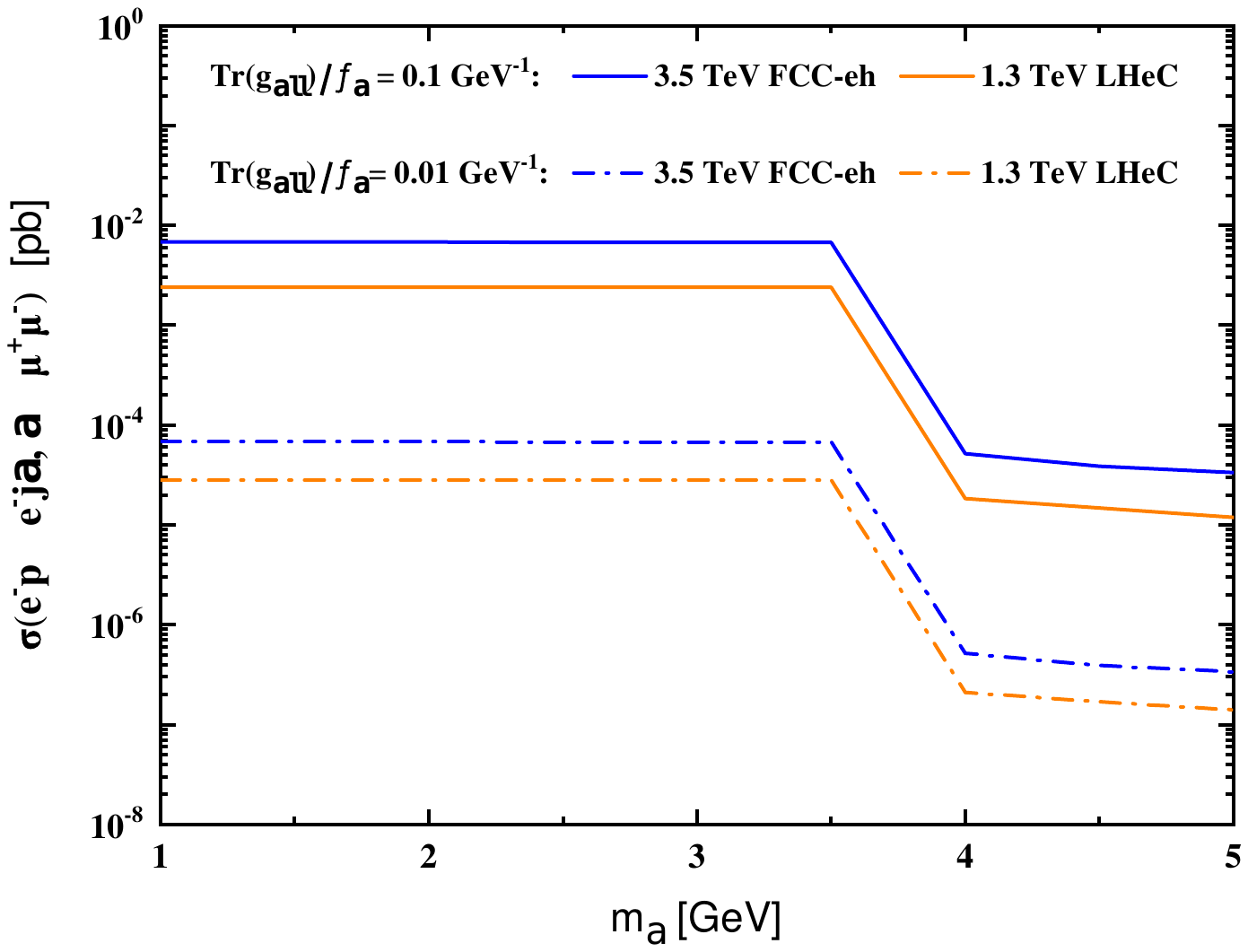}}
\hspace{-0.06\textwidth}
\subfigure[]{\includegraphics [width=0.52\textwidth] {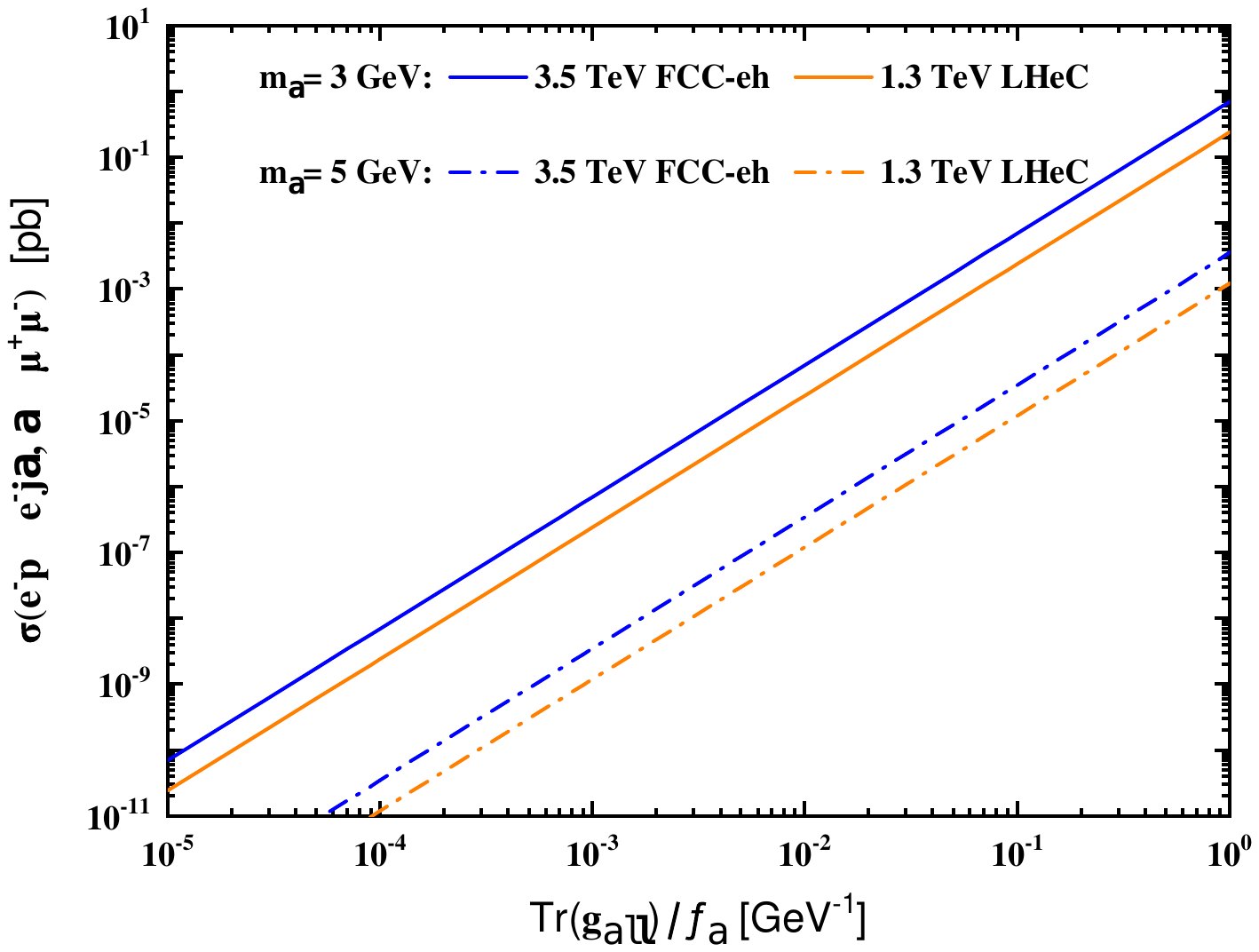}}
\caption{The production cross sections of the signal process $e^{-}p\rightarrow e^-ja(a\rightarrow\mu^{+}\mu^{-})$ as functions of the ALP mass $m_a$~(a) and the ALP-lepton coupling $Tr(g_{a\ell\ell})/f_a$~(b) at the $1.3$ TeV LHeC~(orange) and $3.5$ TeV FCC-eh~(blue) for $c_{R}=0$.}
\label{fig:5}
\end{center}
\end{figure}

\begin{figure}[H]
\begin{center}
\subfigure[]{\includegraphics [scale=0.35] {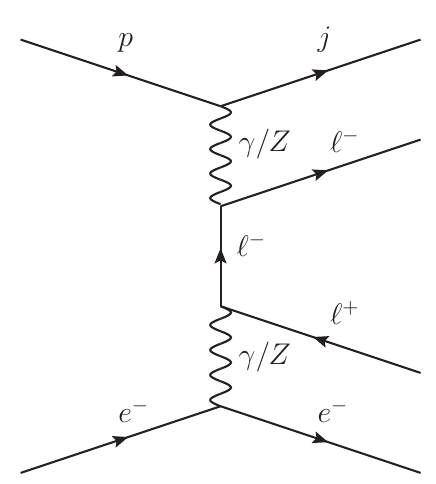}}
\hspace{0.03\textwidth}
\subfigure[]{\includegraphics [scale=0.35] {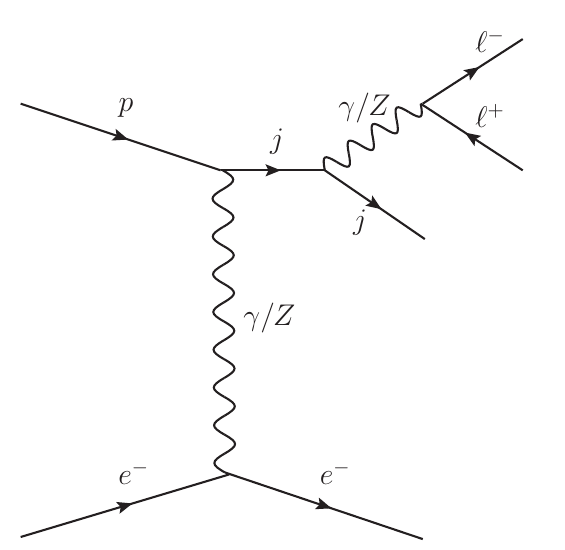}}
\hspace{0.03\textwidth}
\subfigure[]{\includegraphics [scale=0.35] {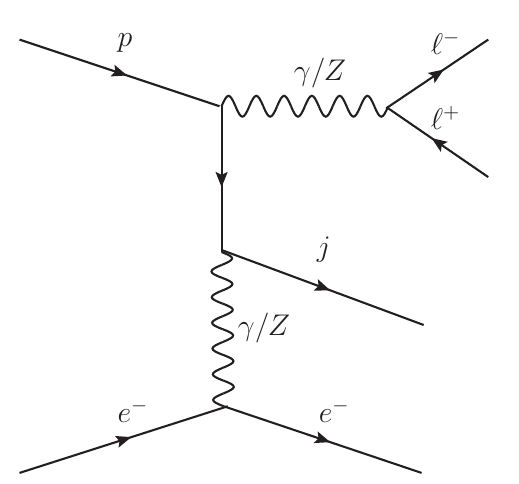}}
\caption{The typical Feynman diagrams for the SM background from the process $e^{-}p\rightarrow e^- j\ell^{+}\ell^{-}~(\ell=e,\mu)$.}
\label{fig:6}
\end{center}
\end{figure}

Now let us begin with the process $e^- p \to e^- j a~(a\rightarrow\mu^{+}\mu^{-})$ at the $1.3$~($3.5$) TeV LHeC~(FCC-eh) with $\mathcal{L}=$ $1$~($2$) ab$^{-1}$ for $c_{R}=0$, in which the ALPs with masses from 1 to 5 GeV are taken into account. The signature of the final state is characterized by the presence of muon pairs, an electron and a light jet $(u$, $d$, $c$, or $s)$. The production cross sections of the signal process at the 1.3~(3.5) TeV LHeC~(FCC-eh) as functions of the ALP mass $m_a$ and the ALP-lepton coupling $Tr(g_{a\ell\ell})/f_a$ are presented in Fig.~\ref{fig:5}, which have been obtained with the basic cuts applied, as described earlier. As illustrated in Fig.~\ref{fig:5}, the production cross sections decrease slowly with increasing ALP mass $m_a$. However, when the ALP mass $m_a$ is around 3.5 GeV, its values drop sharply due to the opening of the $a\rightarrow\tau^{+}\tau^{-}$ decay channel. The production cross section at the $3.5$ TeV FCC-eh is larger than that at the $1.3$ TeV LHeC due to the higher center-of-mass energy of the FCC-eh, which provides more phase space for ALP production and reduces kinematic suppression. For the ALPs with masses in the range of $1$ $\sim$ $5$ GeV, the values of the signal cross sections range from $2.410 \times 10^{-3}~(6.830 \times 10^{-3})$ to $1.192 \times 10^{-5}~(3.355 \times 10^{-5})$ pb for $Tr(g_{a\ell\ell})/f_a$ equaling to $0.1$ GeV$^{-1}$, while for a weaker coupling of $0.01$ GeV$^{-1}$, they decrease by two orders of magnitude, spanning from $2.810 \times 10^{-5}~(6.810 \times 10^{-5})$ to $1.394 \times 10^{-7}~(3.370 \times 10^{-7})$ pb at the $1.3~(3.5)$ TeV LHeC~(FCC-eh).
When accounting for the uncertainty associated with PDF, the signal cross section varies within $\pm 1\%$ $\sim \pm2\%$. While considering the uncertainty arising from the choices of factorization scale, the signal cross section shows a variation of $\pm 3\%$ $\sim$ $\pm 5\%$. It is evident that these uncertainties have a minimal impact on the signal cross sections. The corresponding SM background considered in our numerical analysis mainly comes from the process $e^{-}p\rightarrow e^- j\ell^{+}\ell^{-}~(\ell=e,\mu)$. The typical Feynman diagrams are shown in Fig.~\ref{fig:6}. The signal cross sections in the considered parameter region are smaller than the background cross sections~$0.041~(0.095)$ pb.

\begin{figure}[H]
\begin{center}
\subfigure[]{\includegraphics [scale=0.36] {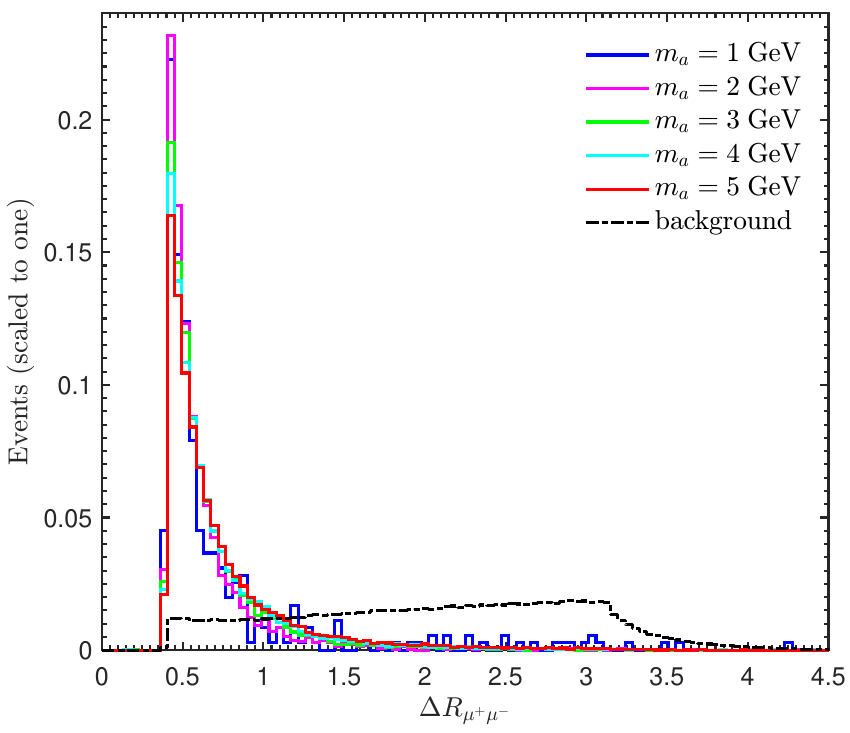}}
\hspace{0.3in}
\subfigure[]{\includegraphics [scale=0.36] {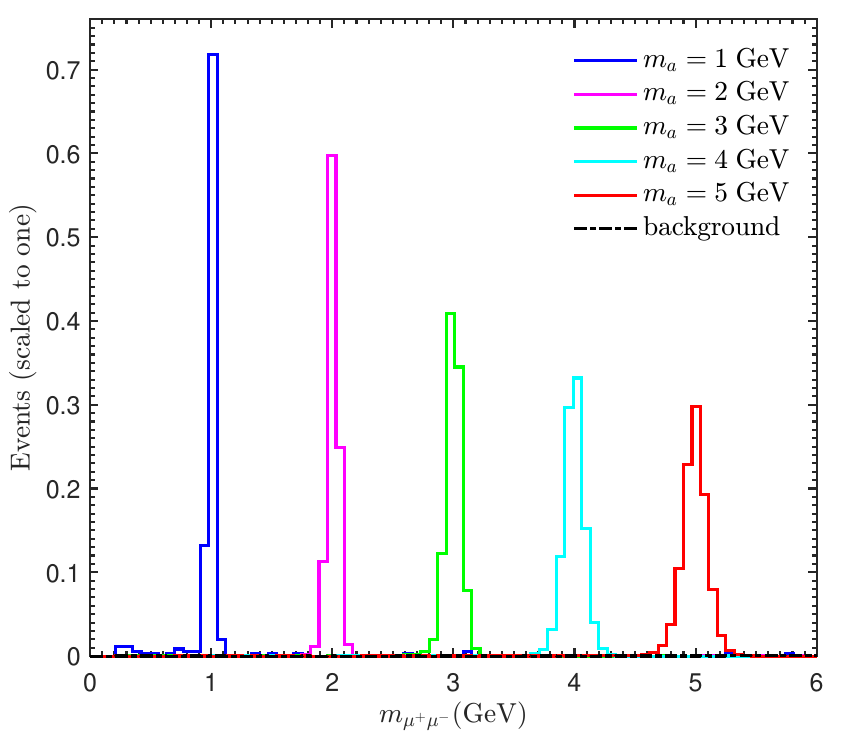}}
\hspace{0.3in}
\subfigure[]{\includegraphics [scale=0.36] {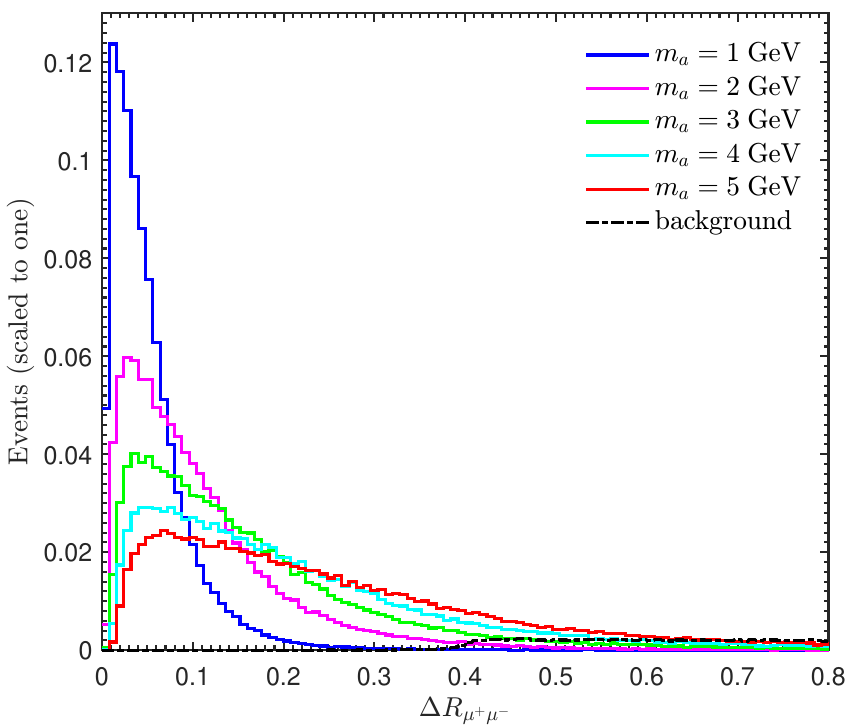}}
\hspace{0.3in}
\subfigure[]{\includegraphics [scale=0.36] {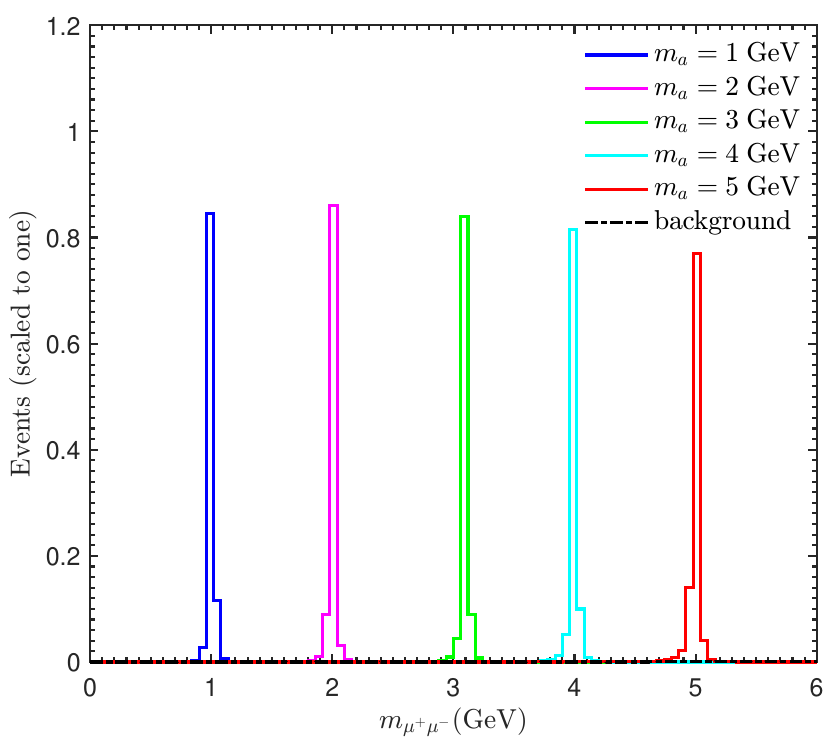}}
\caption{{The normalized distributions of the observables $\Delta R _{\mu^{+}\mu^{-}}$ and $m_{\mu^{+}\mu^{-}}$ for the signal of selected ALP-mass benchmark points and SM background at the 1.3 TeV LHeC with $\mathcal{L}=$ $1$ ab$^{-1}$~(a, b) and 3.5 TeV FCC-eh with $\mathcal{L}=$ $2$ ab$^{-1}$~(c, d).}}
\label{fig:7}
\end{center}
\end{figure}

For the ALPs with masses in the range of $1 \sim 5$ GeV, observables of the angular separation between positive and negative muons $\Delta R _{\mu^{+}\mu^{-}}$ and the invariant mass of a pair of muons $m_{\mu^{+}\mu^{-}}$ are taken to be analyzed. We present in Fig.~\ref{fig:7} the distributions of $\Delta R _{\mu^{+}\mu^{-}}$ and $m_{\mu^{+}\mu^{-}}$ for the signal and background events with typical points of $m_a = 1$, $2$, $3$, $4$, $5$ GeV at the 1.3 TeV LHeC with $\mathcal{L}=$ $1$ ab$^{-1}$~(a, b) and 3.5 TeV FCC-eh with $\mathcal{L}=$ $2$ ab$^{-1}$~(c, d), where the background events include events involving electron and muon. As shown in Fig.~\ref{fig:7}, the two muons from the light ALP decays are typically very close, due to the relatively low momentum of the ALP decay products, resulting in a smaller angular separation between the muons in signal events. In contrast, the background processes often produce lepton pairs with a wider angular separation due to more complex interaction, such as virtual photon exchange, leading to a broader angular distribution. The invariant mass distribution $m_{\mu^{+}\mu^{-}}$ for the signal events has peak around the ALP mass, which becomes wider as the ALP mass increases. While, the SM background events do not show a distinct peak and have a more spread-out invariant mass distribution.

According to the information of these kinematic distributions, improved cuts presented in Table~\ref{tab1} are further imposed for separating the signal from SM background at the $1.3~(3.5)$ TeV LHeC~(FCC-eh), where the particle numbers in the final states are subject to the conditions of $N_{\mu^{+}}\geq1$, $N_{\mu^{-}}\geq1$ and $N_j\geq1$. In Tables~\ref{tab2} and~\ref{tab3}, we show the production cross sections of the signal and background at the LHeC~(FCC-eh) with $\sqrt{s}=1.3~(3.5)$ TeV and $\mathcal{L}=$ $1~(2)$ ab$^{-1}$ after taking the step-by-step cuts for few ALP mass benchmark points and the parameter $Tr(g_{a\nu\nu})/f_a= Tr(g_{a\ell\ell})/f_a=0.1$ GeV$^{-1}$. It can be seen that the background is effectively suppressed. We further calculate the statistical significance $SS=S/\sqrt{S + B}$, where $S$ and $B$ are the number of events for the signal and background, respectively. Large $SS$ values can be attained over a wide range of the parameter space, as illustrated in these tables. The values of $SS$ can range from $2.269~(111.106)$ to $0.666~(4.865)$ for $m_a=1$ $\sim$ $5$ GeV.

\begin{table}[H]
\begin{center}
\setlength{\tabcolsep}{1.5mm}{
\caption{The improved cuts on the signal and background for $1$ GeV $\leq$ $m_a\leq5$ GeV at the $1.3~(3.5)$ TeV LHeC~(FCC-eh).}
\label{tab1}
\resizebox{16cm}{!}{
\begin{tabular}
[c]{c|c c}\hline \hline
\multirow{2}{*}{Cuts}    & \multicolumn{2}{c}{$1$ GeV $\leq$ $m_a\leq5$ GeV }   \\
\cline{2-3}
	       &~~~~~~~LHeC~($\sqrt{s}=1.3$ TeV)~~~   &FCC-eh~($\sqrt{s}=3.5$ TeV)~~~~~~~         \\ \hline
	Cut 1: the particle numbers in the final states  &  $N_{\mu^{+}}\geq1$, $N_{\mu^{-}}\geq1$, $N_j\geq1$ &  $N_{\mu^{+}}\geq1$, $N_{\mu^{-}}\geq1$, $N_j\geq1$  \\
	Cut 2: the angular separation between muons   &  $\Delta R _{\mu^{+}\mu^{-}}<1.1$   &  $\Delta R _{\mu^{+}\mu^{-}}<0.6$\\
	Cut 3: the invariant-mass window cut on the pair of muons        & $|m_{\mu^{+}\mu^{-}}-m_a|\leq 0.2$ GeV & $|m_{\mu^{+}\mu^{-}}-m_a|\leq 0.2$ GeV    \\ \hline \hline
\end{tabular}}}
\end{center}
\end{table}

\begin{table}[H]\tiny
	\centering{
\caption{The production cross sections of the signal and SM background after the improved cuts applied at the $1.3$ TeV LHeC with $\mathcal{L}=$ $1$ ab$^{-1}$ for $Tr(g_{a\nu\nu})/f_a= Tr(g_{a\ell\ell})/f_a=0.1$ GeV$^{-1}$ with the benchmark points $m_a = 1$, $2$, $3$, $4$, $5$ GeV.$~$\label{tab2}}
		\newcolumntype{C}[1]{>{\centering\let\newline\\\arraybackslash\hspace{50pt}}m{#1}}
		\begin{tabular}{m{1.5cm}<{\centering}|m{2.3cm}<{\centering} m{2.3cm}<{\centering} m{2.3cm}<{\centering}  m{2.3cm}<{\centering} m{2.3cm}<{\centering}}
			\hline \hline
      \multirow{2}{*}{Cuts} & \multicolumn{5}{c}{cross sections for signal (background) [pb]}\\
     \cline{2-6}
     & $m_a=1$ GeV  & $m_a=2$ GeV  & $m_a=3$ GeV  & $m_a=4$ GeV & $m_a=5$ GeV  \\ \hline
     Basic Cuts  & \makecell{$2.407\times10^{-3}$\\$(0.041)$} & \makecell{$2.406\times10^{-3}$\\$(0.041)$} &\makecell{$2.397\times10^{-3}$\\$(0.041)$} &\makecell{$1.828\times10^{-5}$\\$(0.041)$} & \makecell{$1.192\times10^{-5}$\\$(0.041)$}
        \\
     Cut 1  & \makecell{$6.680\times10^{-6}$\\$(0.019)$} & \makecell{$8.484\times10^{-5}$\\$(0.019)$} &\makecell{$2.285\times10^{-4}$\\$(0.019)$} &\makecell{$3.140\times10^{-6}$\\$(0.019)$} & \makecell{$3.000\times10^{-6}$\\$(0.019)$}
       \\
     Cut 2  & \makecell{$5.810\times10^{-6}$\\$(3.373\times10^{-3})$} & \makecell{$7.994\times10^{-5}$\\$(3.373\times10^{-3})$} &\makecell{$2.091\times10^{-4}$\\$(3.373\times10^{-3})$} &\makecell{$2.810\times10^{-6}$\\$(3.373\times10^{-3})$} & \makecell{$2.640\times10^{-6}$\\$(3.373\times10^{-3})$}
       \\
     Cut 3  & \makecell{$5.530\times10^{-6}$\\$(4.090\times10^{-7})$} & \makecell{$7.947\times10^{-5}$\\$(1.640\times10^{-7})$} &\makecell{$2.063\times10^{-4}$\\$(2.460\times10^{-7})$} &\makecell{$2.720\times10^{-6}$\\$(8.190\times10^{-7})$} & \makecell{$2.500\times10^{-6}$\\$(1.146\times10^{-5})$}
     \\ \hline
     $SS$  & $2.269$ & $8.905$ & $14.355$ & $1.447$ & $0.666$ \\ \hline \hline
	\end{tabular}}	
\end{table}

\begin{table}[H]\tiny
	\centering{
\caption{Same as Table~\ref{tab2} but for the $3.5$ TeV FCC-eh with $\mathcal{L}=$ $2$ ab$^{-1}$.$~$\label{tab3}}
		\newcolumntype{C}[1]{>{\centering\let\newline\\\arraybackslash\hspace{50pt}}m{#1}}
		\begin{tabular}{m{1.5cm}<{\centering}|m{2.3cm}<{\centering} m{2.3cm}<{\centering} m{2.3cm}<{\centering}  m{2.3cm}<{\centering} m{2.3cm}<{\centering}}
			\hline \hline
      \multirow{2}{*}{Cuts} & \multicolumn{5}{c}{cross sections for signal (background) [pb]}\\
     \cline{2-6}
     & $m_a=1$ GeV  & $m_a=2$ GeV  & $m_a=3$ GeV  & $m_a=4$ GeV & $m_a=5$ GeV  \\ \hline
     Basic Cuts  & \makecell{$6.792\times10^{-3}$\\$(0.095)$} & \makecell{$6.779\times10^{-3}$\\$(0.095)$} &\makecell{$6.771\times10^{-3}$\\$(0.095)$} &\makecell{$5.167\times10^{-5}$\\$(0.095)$} & \makecell{$3.372\times10^{-5}$\\$(0.095)$}
        \\
     Cut 1  & \makecell{$6.182\times10^{-3}$\\$(0.045)$} & \makecell{$6.013\times10^{-3}$\\$(0.045)$} &\makecell{$5.938\times10^{-3}$\\$(0.045)$} &\makecell{$4.520\times10^{-5}$\\$(0.045)$} & \makecell{$2.941\times10^{-5}$\\$(0.045)$}
       \\
     Cut 2  & \makecell{$6.180\times10^{-3}$\\$(2.457\times10^{-3})$} & \makecell{$5.988\times10^{-3}$\\$(2.457\times10^{-3})$} &\makecell{$5.807\times10^{-3}$\\$(2.457\times10^{-3})$} &\makecell{$4.286\times10^{-5}$\\$(2.457\times10^{-3})$} & \makecell{$2.677\times10^{-5}$\\$(2.457\times10^{-3})$}
       \\
     Cut 3  & \makecell{$6.173\times10^{-3}$\\$(1.140\times10^{-6})$} & \makecell{$5.973\times10^{-3}$\\$(9.400\times10^{-7})$} &\makecell{$5.782\times10^{-3}$\\$(7.600\times10^{-7})$} &\makecell{$4.257\times10^{-5}$\\$(7.600\times10^{-7})$} & \makecell{$2.647\times10^{-5}$\\$(3.273\times10^{-5})$}
     \\ \hline
     $SS$  & $111.106$ & $109.285$ & $107.528$ & $9.146$ & $4.865$ \\ \hline \hline
	\end{tabular}}	
\end{table}

\begin{figure}[H]
\begin{center}
\centering\includegraphics [scale=0.32] {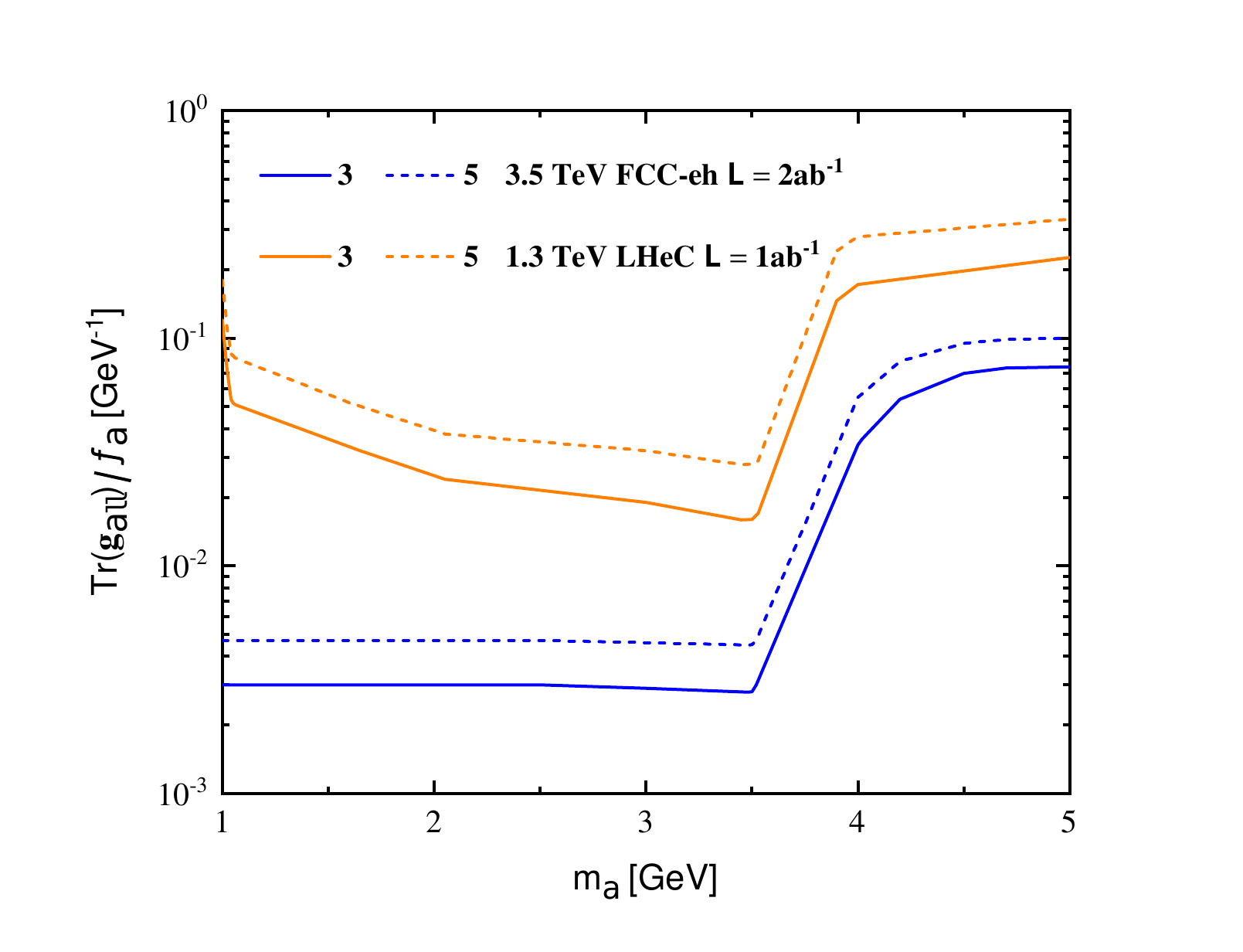}
\caption{The $3\sigma$ and $5\sigma$ curves in the $m_a - Tr(g_{a\ell\ell})/f_a$ plane for the three-lepton final state process $e^- p \to e^- j a~(a \to \mu^+ \mu^-)$ at the $1.3$~($3.5$) TeV LHeC~(FCC-eh) with $\mathcal{L}=$ $1$~($2$) ab$^{-1}$ for $c_{R}=0$.}
\label{fig:8}
\end{center}
\end{figure}

In Fig.~\ref{fig:8}, we plot the $3\sigma$ and $5\sigma$ curves in the plane of $m_a - Tr(g_{a\ell\ell})/f_a$ for the $1.3$ TeV LHeC with $\mathcal{L}=$ $1$ ab$^{-1}$~(orange solid and dashed lines) and $3.5$ TeV FCC-eh with $\mathcal{L}=$ $2$ ab$^{-1}$~(blue solid and dashed lines). As shown in Fig.~\ref{fig:8}, we can obtain that the prospective sensitivities respectively as $1.590\times10^{-2}$~($2.780\times10^{-3}$) GeV$^{-1}$ $\leq$ $Tr(g_{a\ell\ell})/f_a$ $\leq$ $0.226$~($7.500\times10^{-2}$) GeV$^{-1}$ and $2.780\times10^{-2}$~($4.480\times10^{-3}$) GeV$^{-1}$ $\leq$ $Tr(g_{a\ell\ell})/f_a$ $\leq$ $0.334$~($0.100$) GeV$^{-1}$ at $3\sigma$ and $5\sigma$ confidence levels~(CL) for the ALPs with masses in the range of $1$ $\sim$ $5$ GeV. It is obvious that the $3.5$ TeV FCC-eh might have greater potential than the $1.3$ TeV LHeC for detecting the ALP-lepton couplings via the three-lepton final state process $e^{-}p\rightarrow e^-ja(a\rightarrow\mu^{+}\mu^{-})$ within the ALP mass range considered in this paper.

\subsubsection{The process $e^{-}p\rightarrow e^-ja~(a \to \tau^+ \tau^-, \tau^+ \tau^- \to \ell^+ \ell^{-} \slashed E, \ell= e,\mu)$} \label{subsec:second}

\begin{figure}[H]
\begin{center}
\subfigure[]{\includegraphics [width=0.46\textwidth] {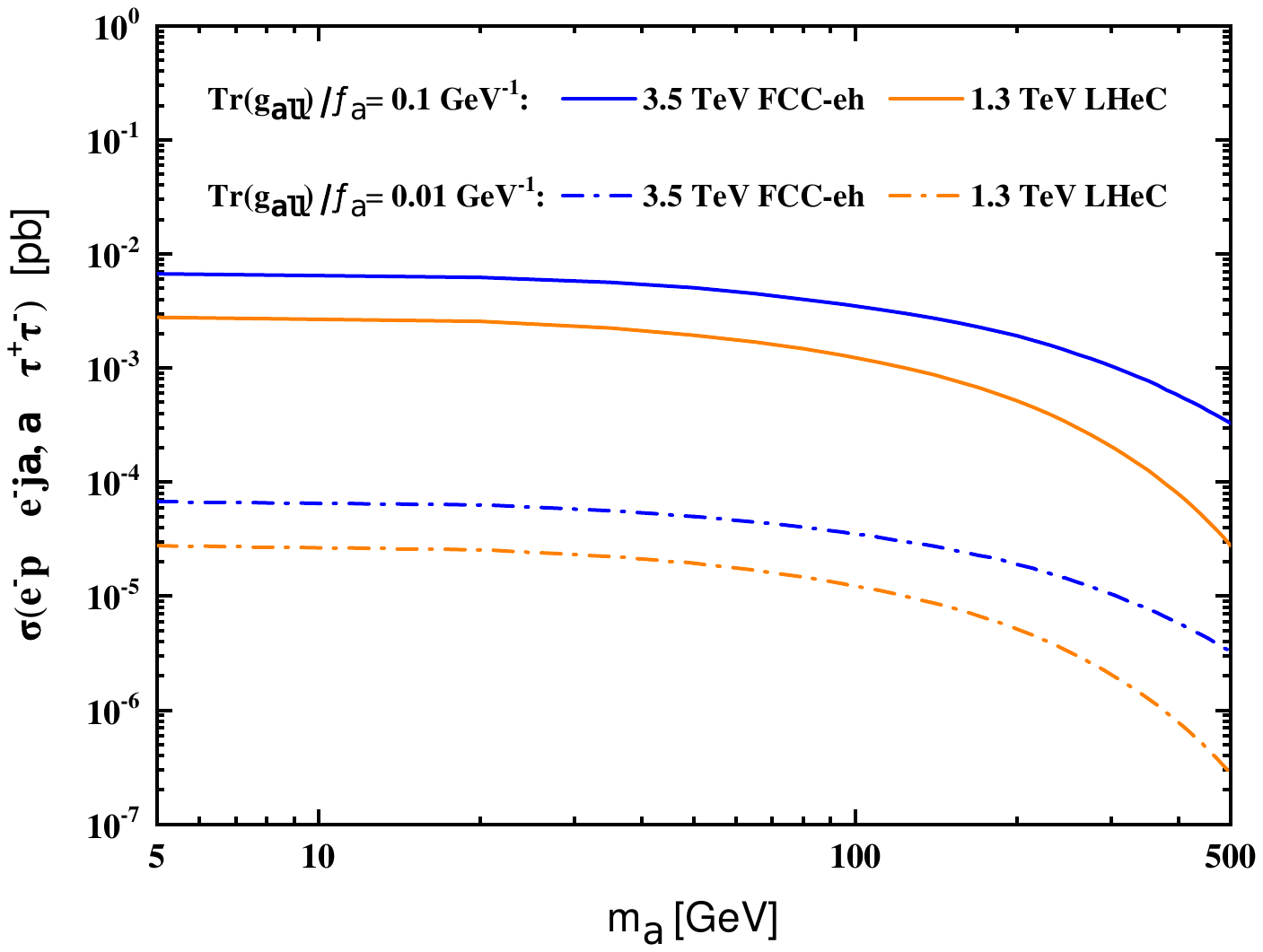}}
\hspace{-0.03\textwidth}
\subfigure[]{\includegraphics [width=0.46\textwidth] {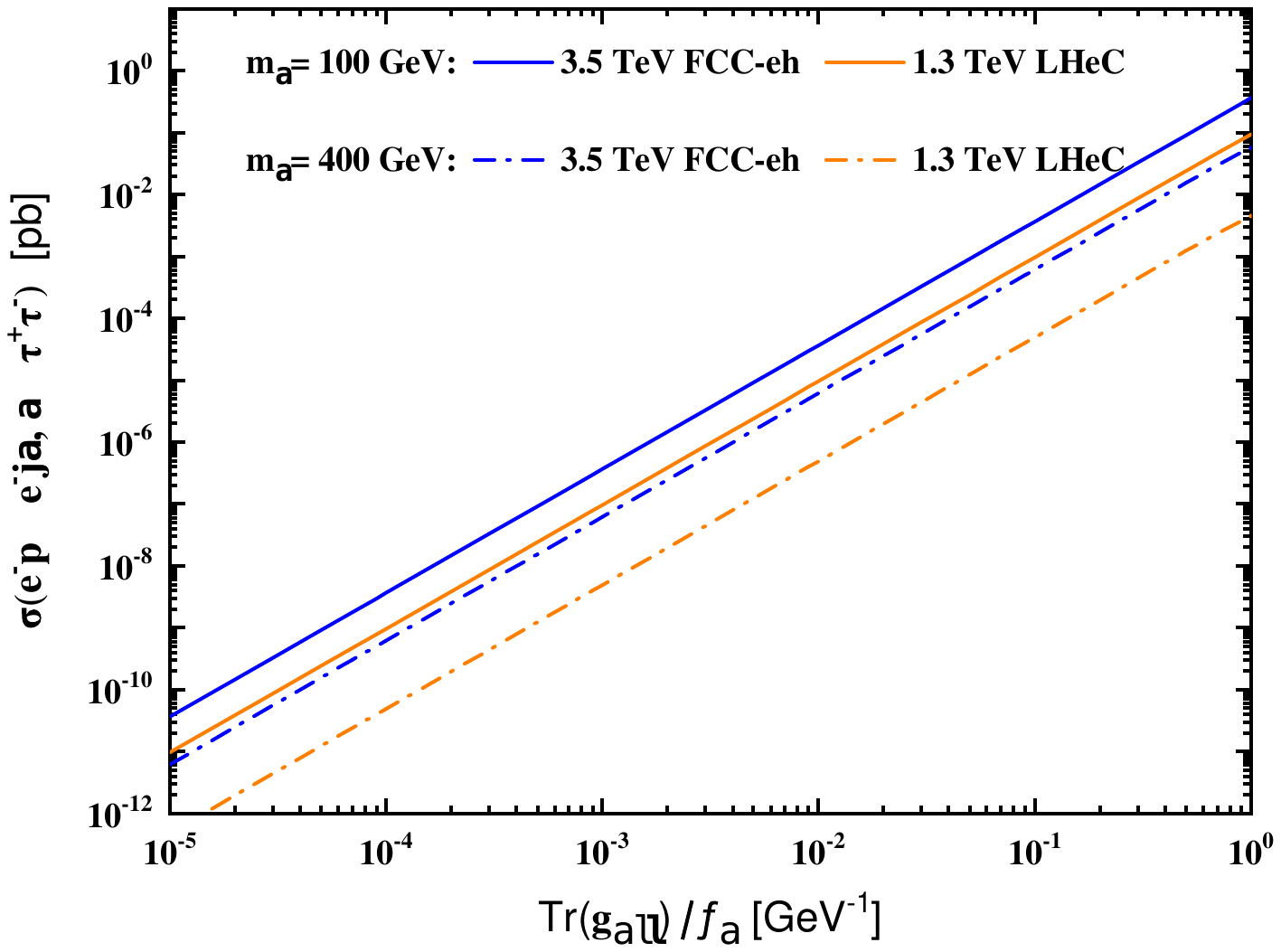}}
\caption{The production cross sections of the signal process $e^{-}p\rightarrow e^-ja(a\rightarrow\tau^{+}\tau^{-})$ as functions of the ALP mass $m_a$~(a) and the ALP-lepton coupling $Tr(g_{a\ell\ell})/f_a$~(b) at the $1.3$ TeV LHeC~(orange) and $3.5$ TeV FCC-eh~(blue) for $c_{R}=0$.}
\label{fig:9}
\end{center}
\end{figure}

In this subsection, we explore the projected sensitivity to the ALP-lepton coupling $Tr(g_{a\ell\ell})/f_a$ via three-lepton final state process $e^{-}p\rightarrow e^-ja~(a \to \tau^+ \tau^-)$ with the sequential leptonic decays of taus at the $1.3$~($3.5$) TeV LHeC~(FCC-eh) with $\mathcal{L}=$ $1$~($2$) ab$^{-1}$ for $c_{R}=0$, in which the ALPs with masses in the range of $5 \sim 500$ GeV are taken into account. The value of the branching ratio $Br(\tau^+ \tau^- \to \ell^+ \ell^{-} \slashed E)$ is approximately 12.397$\%$, which is estimated from $Br(\tau \to \mu \nu\nu)\approx 17.39\%$ and $Br(\tau \to e \nu\nu)\approx 17.82\%$~\cite{ParticleDataGroup:2024cfk}. The signature of the final state is characterized by the presence of invisible neutrinos and three leptons plus a light jet ($u$, $d$, $c$, or $s$). The production cross sections of the signal process $e^{-}p\rightarrow e^-ja~(a \to \tau^+ \tau^-)$ at the $1.3~(3.5)$ TeV LHeC~(FCC-eh) as functions of the ALP mass $m_a$ and the ALP-lepton coupling $Tr(g_{a\ell\ell})/f_a$ are shown in Fig.~\ref{fig:9}, which have been obtained with the basic cuts applied. For the ALPs with masses ranging from $5$ to $500$ GeV, the production cross sections are in the ranges of $2.780 \times 10^{-3}~(6.700 \times 10^{-3})$ $\sim$ $2.772 \times 10^{-5}~(3.270 \times 10^{-4})$ pb and $2.77 \times 10^{-5}~(6.710 \times 10^{-5})$ $\sim$ $2.780 \times 10^{-7}~(3.260 \times 10^{-6})$ pb for $Tr(g_{a\ell\ell})/f_a$ equaling to $0.1$ GeV$^{-1}$ and $0.01$ GeV$^{-1}$, respectively.
When accounting for the uncertainty associated with PDF, the signal cross section varies within $\pm 1\%$ $\sim \pm3\%$. While considering the uncertainty arising from the choices of factorization scale, the signal cross section shows a variation of $\pm 5\%$ $\sim$ $\pm 10\%$. It is clear that these uncertainties have a small effect on the signal cross sections.

\vspace{-12pt}

\begin{figure}[H]
\begin{center}
\subfigure[]{\includegraphics [scale=0.32] {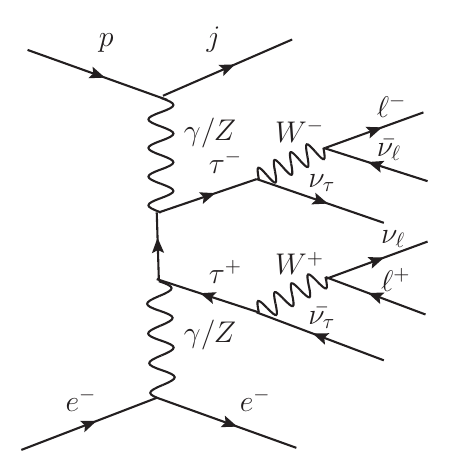}}
\subfigure[]{\includegraphics [scale=0.32] {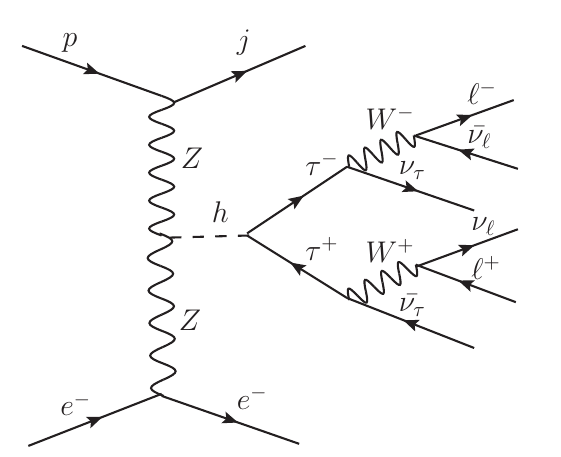}}
\subfigure[]{\includegraphics [scale=0.32] {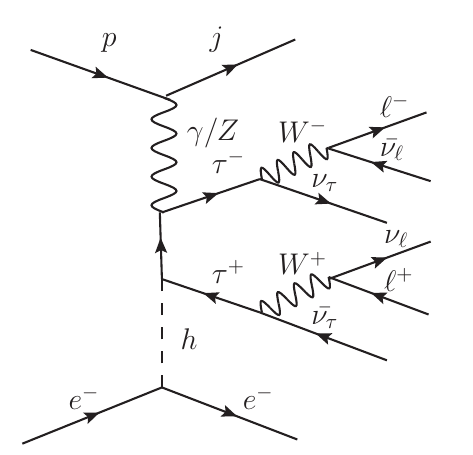}}
\subfigure[]{\includegraphics [scale=0.32] {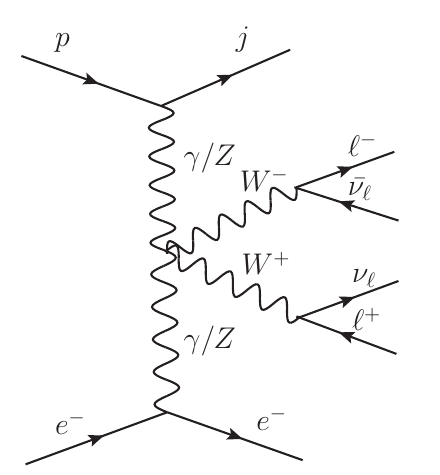}}
\subfigure[]{\includegraphics [scale=0.32] {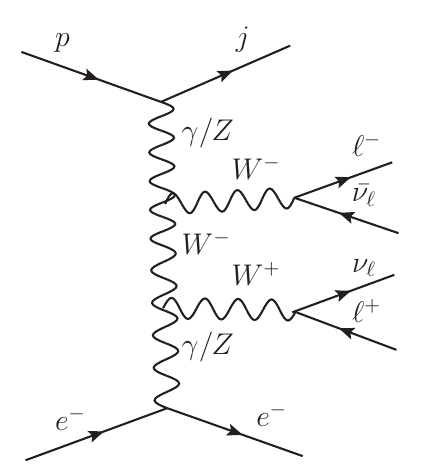}}
\subfigure[]{\includegraphics [scale=0.32] {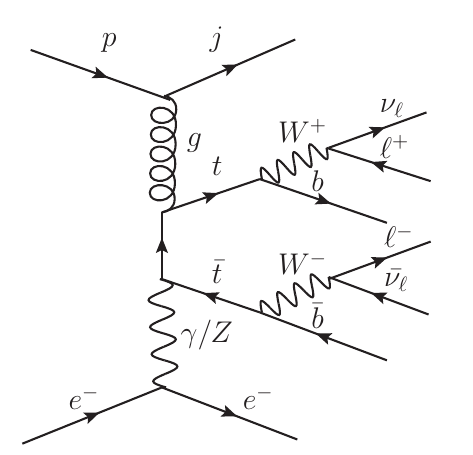}}
\caption{The typical Feynman diagrams for the SM background from the processes $e^- p \to e^- j \tau^+ \tau^-~(\tau^+ \tau^- \to \ell^+ \ell^- \slashed E$)~(a,b,c), $e^- p \to e^- j W^+ W^-~(W^+ W^- \to \ell^+ \ell^- \slashed E$)~(d,e) and $e^- p \to e^- j t \bar{t}~(t \bar{t} \to \ell^+ \ell^- b \bar{b} \slashed E)$~(f).}
\label{fig:10}
\end{center}
\end{figure}

The main irreducible SM background arises from the processes $e^- p \to e^- j \tau^+ \tau^-$ ($\tau^+ \tau^- \to \ell^+ \ell^- \slashed E$, $\ell = e, \mu$), as shown in Fig.~\ref{fig:10}(a,b,c). Subdominant contributions come from the processes $e^- p \to e^- j W^+ W^-$ ($W^+ W^- \to \ell^+ \ell^- \slashed E$, $\ell = e, \mu$), which are also included and illustrated in Fig.~\ref{fig:10}(d,e). Additionally, the processes $e^- p \to e^- j Z W^+ W^-$ ($Z \to \nu_\ell \bar{\nu}_\ell$, $W^+ W^- \to \ell^+ \ell^- \slashed E$, $\ell = e, \mu$) could contribute to the SM background. However, they can be safely ignored as their contributions to the SM background are estimated to be less than $0.001$~($0.01$)$\%$ for the $1.3$~($3.5$) TeV LHeC~(FCC-eh). Furthermore, the process $e^- p \to e^- j t \bar{t}$, with the top and  antitop quarks decaying as $t \to b \ell^{+} \nu_{\ell}$ and $\bar{t} \to \bar{b} \ell^{-} \bar{\nu}_{\ell}$, as depicted in Fig.~\ref{fig:10}(f), is included as a reducible background in our analysis due to the potential misidentification of $b$-quarks as jets. Although its contribution to the total SM background is relatively small, it is systematically incorporated into our calculations. The cross section of the SM background is $0.011~(0.037)$ pb at the $1.3$~($3.5$) TeV LHeC~(FCC-eh).

\begin{figure}[H]
\begin{center}
\subfigure[]{\includegraphics [scale=0.34] {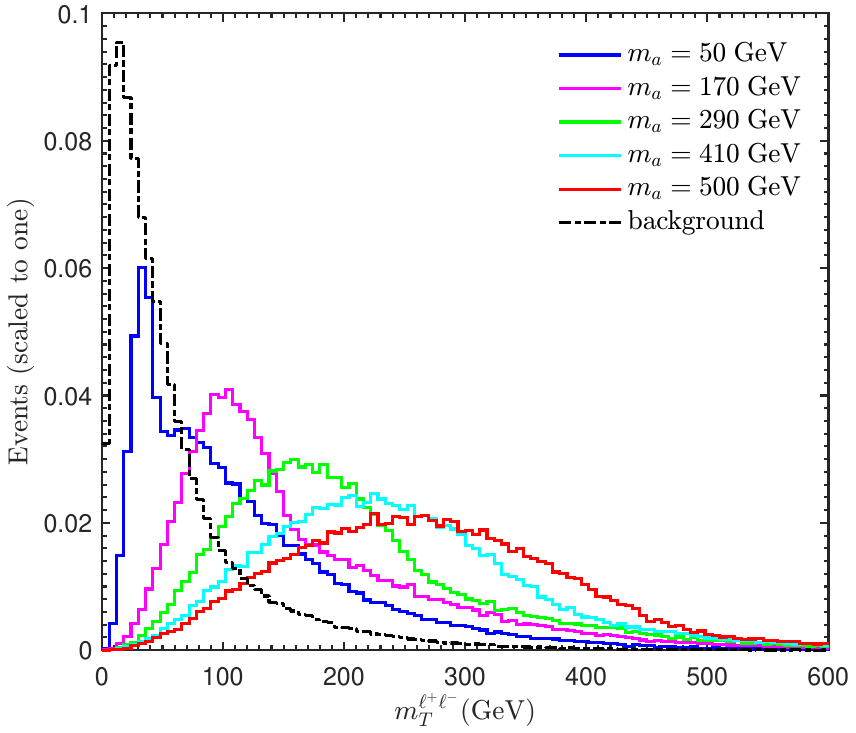}}
\hspace{0.1in}
\subfigure[]{\includegraphics [scale=0.34] {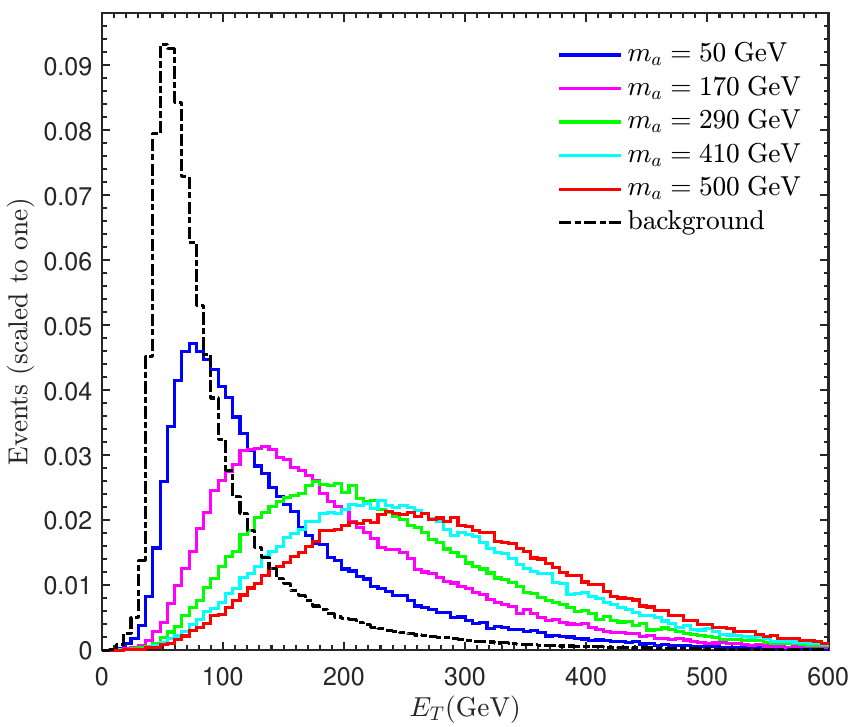}}
\hspace{0.1in}
\subfigure[]{\includegraphics [scale=0.34] {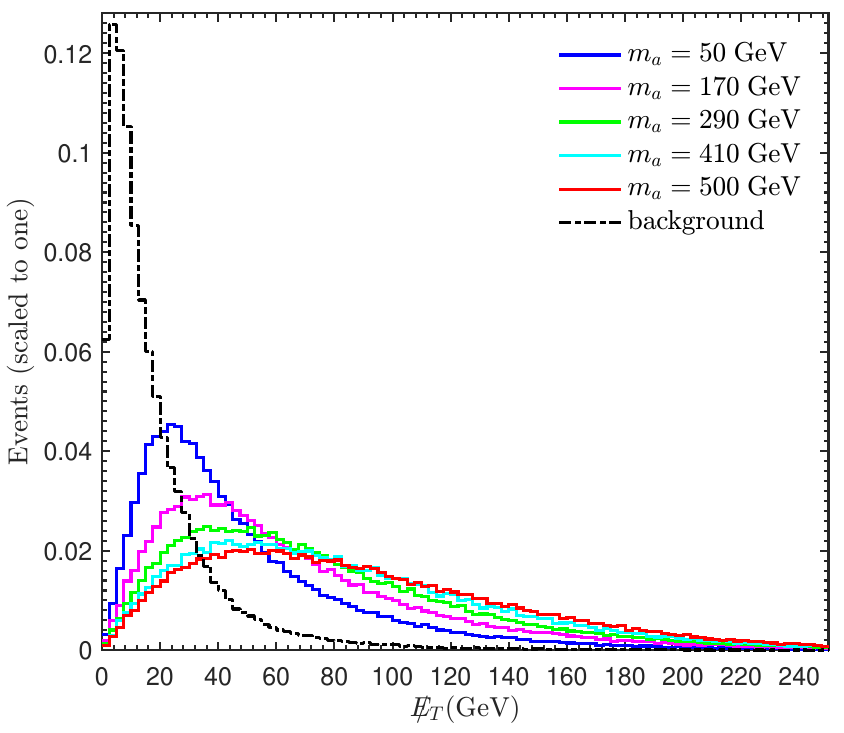}}
\hspace{0.1in}
\subfigure[]{\includegraphics [scale=0.34] {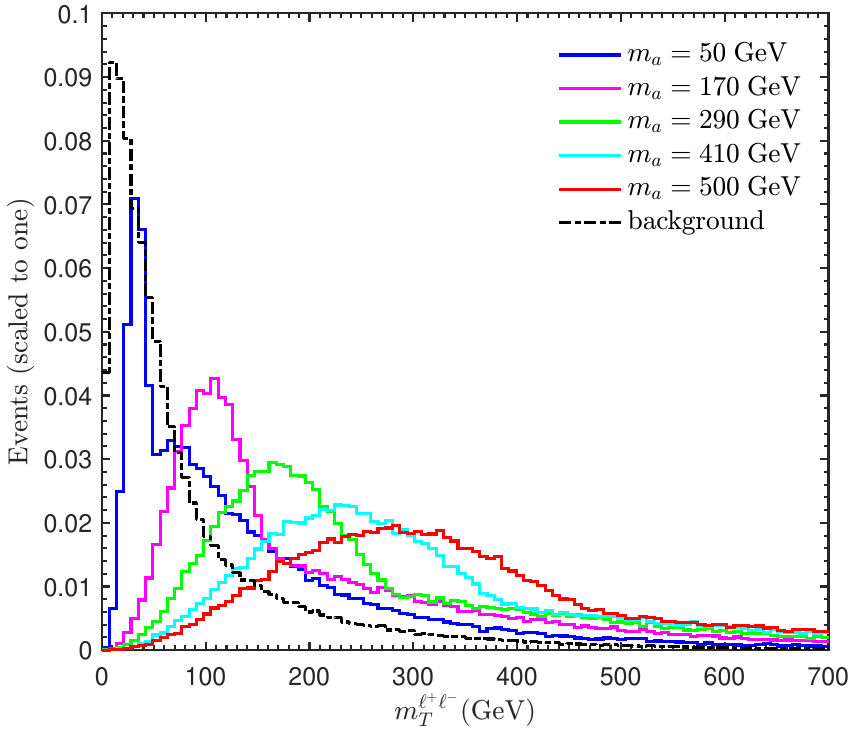}}
\hspace{0.1in}
\subfigure[]{\includegraphics [scale=0.34] {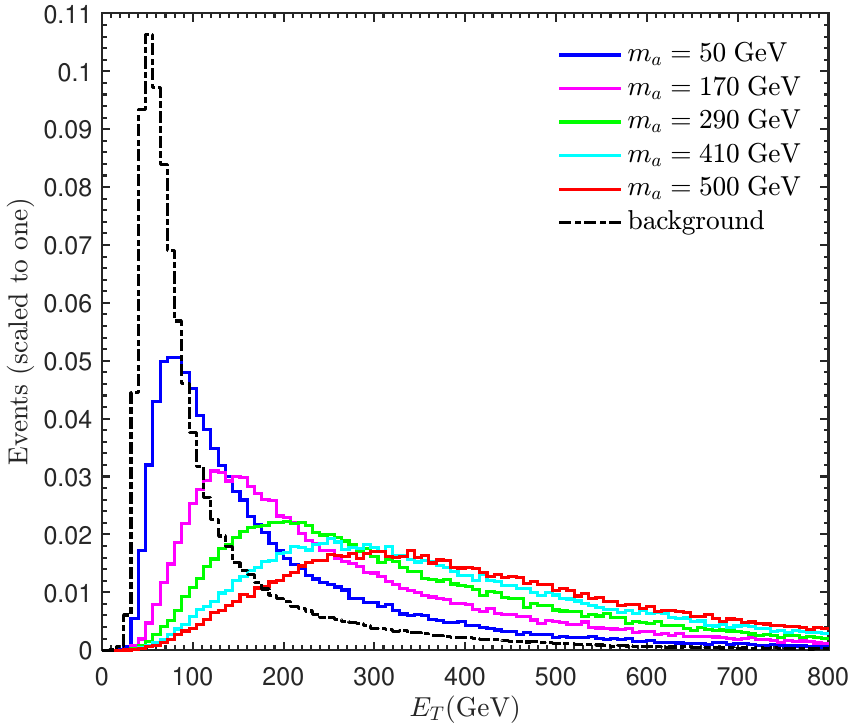}}
\hspace{0.1in}
\subfigure[]{\includegraphics [scale=0.34] {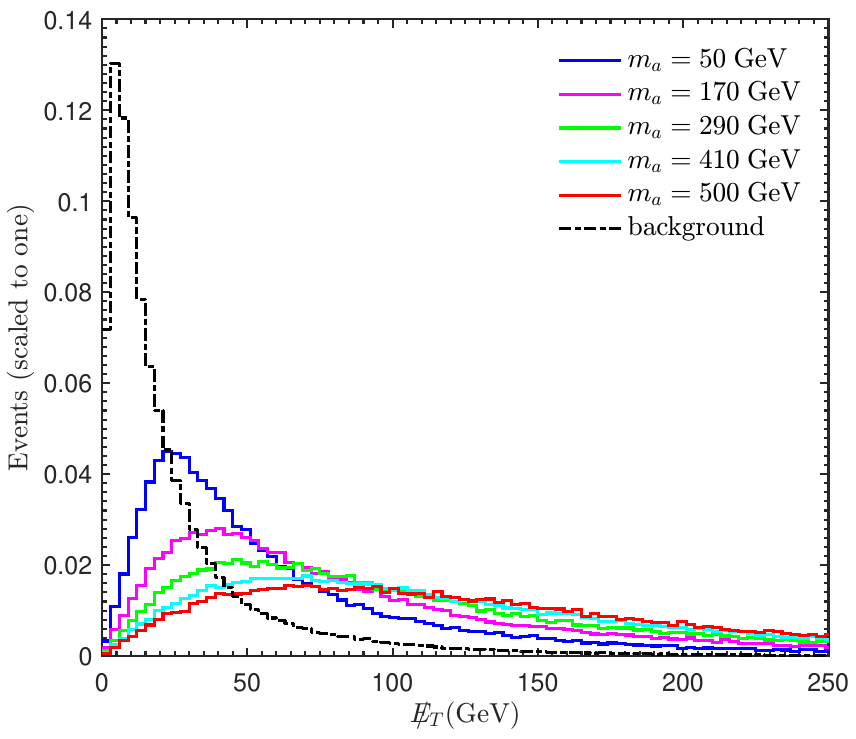}}
\caption{The normalized distributions of the observables $m_T^{\ell^{+}\ell^{-}}$, $E_T$ and $\slashed E_T$ for the signal of selected ALP-mass benchmark points and SM background at the 1.3 TeV LHeC with $\mathcal{L}=$ $1$ ab$^{-1}$~(a, b, c) and 3.5 TeV FCC-eh with $\mathcal{L}=$ $2$ ab$^{-1}$~(d, e, f).}
\label{fig:11}
\end{center}
\end{figure}

Although the signal may be easily overshadowed by the large SM background, there are significant kinematic differences between them that can be exploited to distinguish the signal from the SM background. The transverse mass of the system comprised of the lepton pairs and the missing momentum, $m_T^{\ell^+\ell^-}$, the total transverse energy $E_T$ as well as the missing transverse energy $\slashed E_T$ are taken as the variables for analysis. The normalized distributions of these kinematic variables at the 1.3 TeV LHeC with $\mathcal{L}=$ $1$ ab$^{-1}$~(a, b, c) and 3.5 TeV FCC-eh with $\mathcal{L}=$ $2$ ab$^{-1}$~(d, e, f) are shown in Fig.~\ref{fig:11} with several benchmark ALP mass points $m_a = 50$, $170$, $290$, $410$, $500$ GeV selected for illustration. As shown in Fig.~\ref{fig:11}, the $m_T^{\ell^+\ell^-}$ distribution for the SM background is concentrated in the very low mass range. In contrast, the signal is primarily concentrated in higher mass range. The $E_T$ distributions for the SM background events are concentrated at around $70$ GeV. Additionally, $\slashed E_T$ is a useful kinematic variable for distinguishing the signal from the SM background. For the SM background, the distribution peaks at lower values of $\slashed E_T$, while for the signal, it shifts to relatively higher values. According to the features of these kinematic distributions, different optimized kinematical cuts are applied to reduce background and improve the statistical significance, as listed in Table~\ref{tab4}. The production cross sections of the signal and  background after imposing improved cuts for few ALP mass benchmark points and the parameter $Tr(g_{a\ell\ell})/f_a=Tr(g_{a\nu\nu})/f_a=0.1$ GeV$^{-1}$ at the $1.3~(3.5)$ TeV LHeC~(FCC-eh) with $\mathcal{L}=$ $1~(2)$ ab$^{-1}$ are given in Tables~\ref{tab5} and~\ref{tab6}. The $SS$ values obtained with the selection strategy are summarized in the last row of these tables. The $SS$ values can reach $2.097~(3.172)$ $\sim$ $0.106~(0.728)$ for $m_a=5$ $\sim$ $500$ GeV at the $1.3~(3.5)$ TeV LHeC~(FCC-eh) with $\mathcal{L}=$ $1~(2)$ ab$^{-1}$.

\begin{table}[H]\tiny
\begin{center}
\caption{The improved cuts on the signal and background for $5$ GeV $\leq$ $m_a\leq500$ GeV at the $1.3~(3.5)$ TeV LHeC~(FCC-eh).}
\label{tab4}
\resizebox{15cm}{!}{
\begin{tabular}
[c]{c|c c}\hline \hline
\multirow{2}{*}{Cuts}    & \multicolumn{2}{c}{ $5$ GeV $\leq$ $m_a\leq500$ GeV }   \\
\cline{2-3}
	       &~~~~~~~LHeC~($\sqrt{s}=1.3$ TeV)~~~~~~~ &FCC-eh~($\sqrt{s}=3.5$ TeV)~~~~~~~       \\ \hline
	Cut 1: the particle numbers in the final states     &  $N_{\ell^-}\geq1$,   $N_{\ell^+}\geq1$, $N_j\geq1$  &  $N_{\ell^-}\geq1$,   $N_{\ell^+}\geq1$, $N_j\geq1$\\

	Cut 2: the transverse mass of the system comprised of \\ the lepton pairs and the missing momentum  &  $m_T^{\ell^+ \ell^-}>120$ GeV    &  $m_T^{\ell^+ \ell^-}>140$ GeV \\

    Cut 3: the total transverse energy      & $E_T>140$ GeV   & $E_T>170$ GeV \\

	Cut 4: the missing transverse energy      &  ${E\mkern-10.5 mu/_T}>35$ GeV    &  ${E\mkern-10.5 mu/_T}>45$ GeV \\ \hline \hline
\end{tabular}}
\end{center}
\end{table}

\begin{table}[H]\tiny
	\centering{
\caption{The production cross sections of the signal and SM background after the improved cuts applied for $Tr(g_{a\ell\ell})/f_a=Tr(g_{a\nu\nu})/f_a=0.1$ GeV$^{-1}$ at the $1.3$ TeV LHeC with $\mathcal{L}=$ $1$ ab$^{-1}$ with benchmark points $m_a = 50$, $170$, $290$, $410$, $500$ GeV.$~$\label{tab5}}
		\newcolumntype{C}[1]{>{\centering\let\newline\\\arraybackslash\hspace{50pt}}m{#1}}
		\begin{tabular}{m{1.5cm}<{\centering}|m{2.3cm}<{\centering} m{2.3cm}<{\centering} m{2.3cm}<{\centering}  m{2.3cm}<{\centering} m{2.3cm}}
			\hline \hline
      \multirow{2}{*}{Cuts} & \multicolumn{5}{c}{cross sections for signal (background) [pb]}\\
     \cline{2-6}
     & $m_a=50$ GeV  & $m_a=170$ GeV  & $m_a=290$ GeV & $m_a=410$ GeV &  $m_a=500$ GeV   \\ \hline
     Basic Cuts  & \makecell{$2.319\times10^{-4}$\\$(0.011)$} & \makecell{$8.007\times10^{-5}$\\$(0.011)$} &\makecell{$2.690\times10^{-5}$\\$(0.011)$} &\makecell{$8.411\times10^{-6}$\\$(0.011)$} & \makecell{$3.312\times10^{-6}$\\$(0.011)$}
        \\
     Cut 1  & \makecell{$2.038\times10^{-4}$\\$(8.257\times10^{-3})$} & \makecell{$7.065\times10^{-5}$\\$(8.257\times10^{-3})$} &\makecell{$2.355\times10^{-5}$\\$(8.257\times10^{-3})$} &\makecell{$7.316\times10^{-6}$\\$(8.257\times10^{-3})$} & \makecell{$2.862\times10^{-6}$\\$(8.257\times10^{-3})$}
       \\
     Cut 2  &\makecell{$7.840\times10^{-5}$\\$(1.109\times10^{-3})$}  & \makecell{$4.146\times10^{-5}$\\$(1.109\times10^{-3})$} &\makecell{$1.908\times10^{-5}$\\$(1.109\times10^{-3})$} &\makecell{$6.530\times10^{-6}$\\$(1.109\times10^{-3})$} & \makecell{$2.632\times10^{-6}$\\$(1.109\times10^{-3})$}
       \\
     Cut 3  & \makecell{$5.130\times10^{-5}$\\$(5.836\times10^{-4})$} & \makecell{$3.353\times10^{-5}$\\$(5.836\times10^{-4})$} &\makecell{$1.718\times10^{-5}$\\$(5.836\times10^{-4})$} &\makecell{$6.120\times10^{-6}$\\$(5.836\times10^{-4})$} & \makecell{$2.506\times10^{-6}$\\$(5.836\times10^{-4})$}
     \\
     Cut 4  & \makecell{$4.556\times10^{-5}$\\$(4.264\times10^{-4})$} & \makecell{$2.861\times10^{-5}$\\$(4.264\times10^{-4})$} &\makecell{$1.457\times10^{-5}$\\$(4.264\times10^{-4})$} &\makecell{$5.300\times10^{-6}$\\$(4.264\times10^{-4})$} & \makecell{$2.192\times10^{-6}$\\$(4.264\times10^{-4})$}
     \\
     \hline
     $SS$  & $2.097$ & $1.341$ & $0.694$ & $0.255$ & ~~~~~~~~~~~$0.106$  \\ \hline \hline
	\end{tabular}}	
\end{table}

\begin{table}[H]\tiny
	\centering{
\caption{Same as Table~\ref{tab5} but for the $3.5$ TeV FCC-eh with $\mathcal{L}=$ $2$ ab$^{-1}$.$~$\label{tab6}}
		\newcolumntype{C}[1]{>{\centering\let\newline\\\arraybackslash\hspace{50pt}}m{#1}}
		\begin{tabular}{m{1.5cm}<{\centering}|m{2.3cm}<{\centering} m{2.3cm}<{\centering} m{2.3cm}<{\centering}  m{2.3cm}<{\centering} m{2.3cm}}
			\hline \hline
      \multirow{2}{*}{Cuts} & \multicolumn{5}{c}{cross sections for signal (background) [pb]}\\
     \cline{2-6}
     & $m_a=50$ GeV  & $m_a=170$ GeV  & $m_a=290$ GeV & $m_a=410$ GeV &  $m_a=500$ GeV   \\ \hline
     Basic Cuts  & \makecell{$5.965\times10^{-4}$\\$(0.037)$} & \makecell{$2.718\times10^{-4}$\\$(0.037)$} &\makecell{$1.313\times10^{-4}$\\$(0.037)$} &\makecell{$6.467\times10^{-5}$\\$(0.037)$} & \makecell{$3.876\times10^{-5}$\\$(0.037)$}
        \\
     Cut 1  & \makecell{$5.266\times10^{-4}$\\$(0.029)$} & \makecell{$2.426\times10^{-4}$\\$(0.029)$} &\makecell{$1.164\times10^{-4}$\\$(0.029)$} &\makecell{$5.710\times10^{-5}$\\$(0.029)$} & \makecell{$3.410\times10^{-5}$\\$(0.029)$}
       \\
     Cut 2  &\makecell{$2.039\times10^{-4}$\\$(5.711\times10^{-3})$}  & \makecell{$1.328\times10^{-4}$\\$(5.711\times10^{-3})$} &\makecell{$9.277\times10^{-5}$\\$(5.711\times10^{-3})$} &\makecell{$5.140\times10^{-5}$\\$(5.711\times10^{-3})$} & \makecell{$3.179\times10^{-5}$\\$(5.711\times10^{-3})$}
       \\
     Cut 3  & \makecell{$1.350\times10^{-4}$\\$(3.765\times10^{-3})$} & \makecell{$1.049\times10^{-4}$\\$(3.765\times10^{-3})$} &\makecell{$8.246\times10^{-5}$\\$(3.765\times10^{-3})$} &\makecell{$4.814\times10^{-5}$\\$(3.765\times10^{-3})$} & \makecell{$3.038\times10^{-5}$\\$(3.765\times10^{-3})$}
     \\
     Cut 4  & \makecell{$1.213\times10^{-4}$\\$(2.803\times10^{-3})$} & \makecell{$9.255\times10^{-5}$\\$(2.803\times10^{-3})$} &\makecell{$7.171\times10^{-5}$\\$(2.803\times10^{-3})$} &\makecell{$4.277\times10^{-5}$\\$(2.803\times10^{-3})$} & \makecell{$2.739\times10^{-5}$\\$(2.803\times10^{-3})$}
     \\
     \hline
     $SS$  & $3.172$ & $2.432$ & $1.892$ & $1.134$ & ~~~~~~~~~~~$0.728$  \\ \hline \hline
	\end{tabular}}	
\end{table}

\begin{figure}[H]
\begin{center}
\centering\includegraphics [scale=0.35] {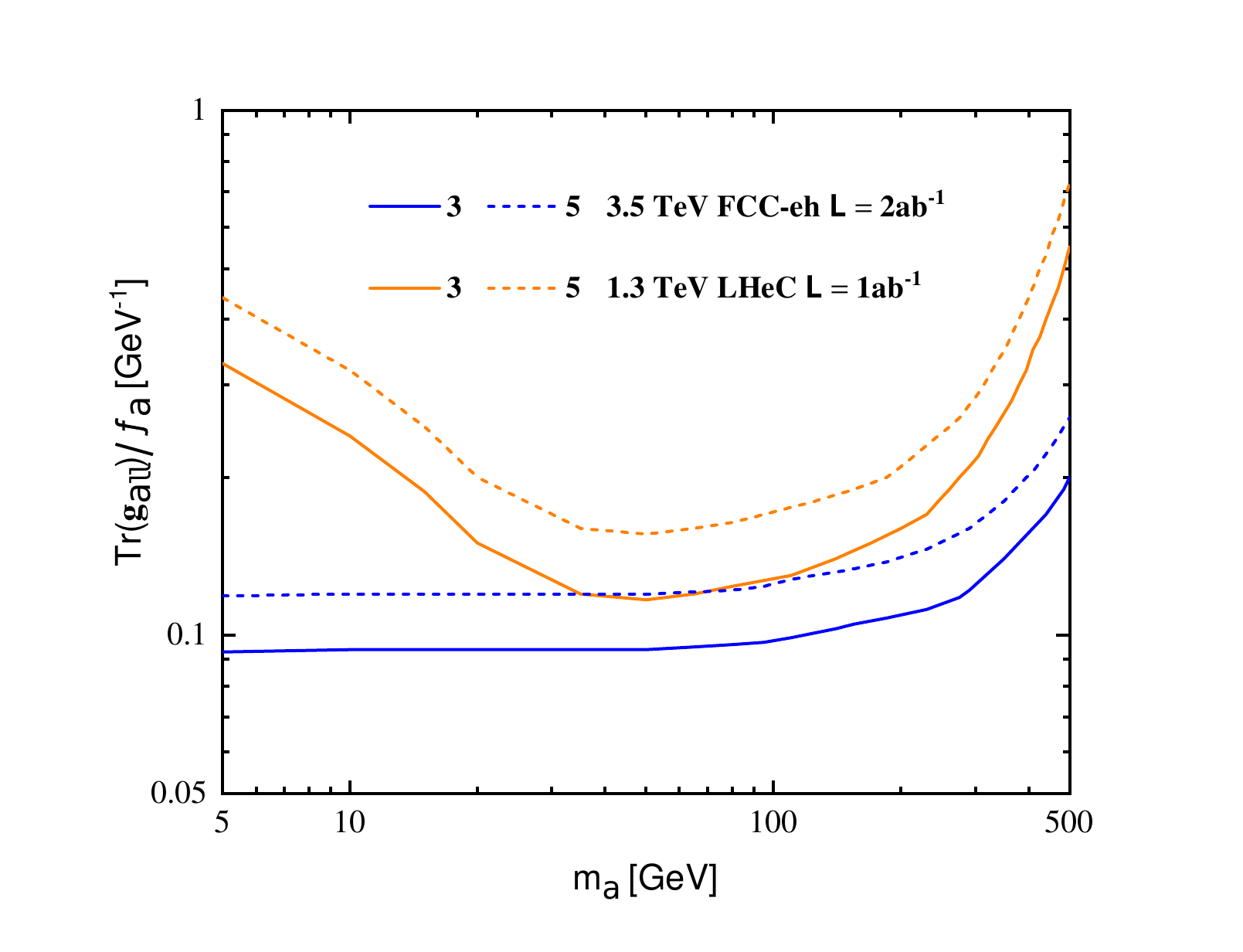}
\caption{The $3\sigma$ and $5\sigma$ curves in the $m_a - Tr(g_{a\ell\ell})/f_a$ plane for the three-lepton final state process $e^{-}p\rightarrow e^-ja~(a \to \tau^+ \tau^-, \tau^+ \tau^- \to \ell^+ \ell^- \slashed E, \ell = e,\mu)$ at the $1.3$~($3.5$) TeV LHeC~(FCC-eh) with $\mathcal{L}=$ $1$~($2$) ab$^{-1}$ for $c_{R}=0$.}
\label{fig:12}
\end{center}
\end{figure}

We plot the $3\sigma$ and $5\sigma$ curves in the plane of $m _a - Tr(g_{a\ell\ell})/f_a$ at the $1.3$ TeV LHeC with $\mathcal{L}=$ $1$ ab$^{-1}$~(orange solid and dashed lines) and $3.5$ TeV FCC-eh with $\mathcal{L}=$ $2$ ab$^{-1}$~(blue solid and dashed lines) in Fig.~\ref{fig:12}. The figure indicates that the projected bounds on the coupling coefficient $Tr(g_{a\ell\ell})/f_a$ can reach $0.117~(0.093)$ $\sim$ $0.550~(0.200)$ GeV$^{-1}$ and $0.156~(0.119)$ $\sim$ $0.730~(0.260)$ GeV$^{-1}$ at $3\sigma$ and $5\sigma$ CL for the ALPs with masses in the range of $5$ $\sim$ $500$ GeV at the LHeC~(FCC-eh) with $\sqrt{s}=1.3~(3.5)$ TeV and $\mathcal{L}=$ $1~(2)$ ab$^{-1}$. Comparing to the projected limits derived in \ref{subsec:first}, we find that the expected limits on the ALP-lepton couplings for the ALPs with masses in the range of $1$ $\sim$ $5$ GeV via the process $e^{-}p \rightarrow e^- j a \, (a \to \mu^+ \mu^-)$ are stricter than those from the process $e^{-}p \rightarrow e^- j a \, (a \to \tau^+ \tau^-, \tau^+ \tau^- \to \ell^+ \ell^- \slashed{E}, \ell = e,\mu)$ for the ALPs with masses in the range of $5$ $\sim$ $500$ GeV at the
LHeC~(FCC-eh).

\begin{figure}[H]
\begin{center}
\centering\includegraphics [scale=0.45] {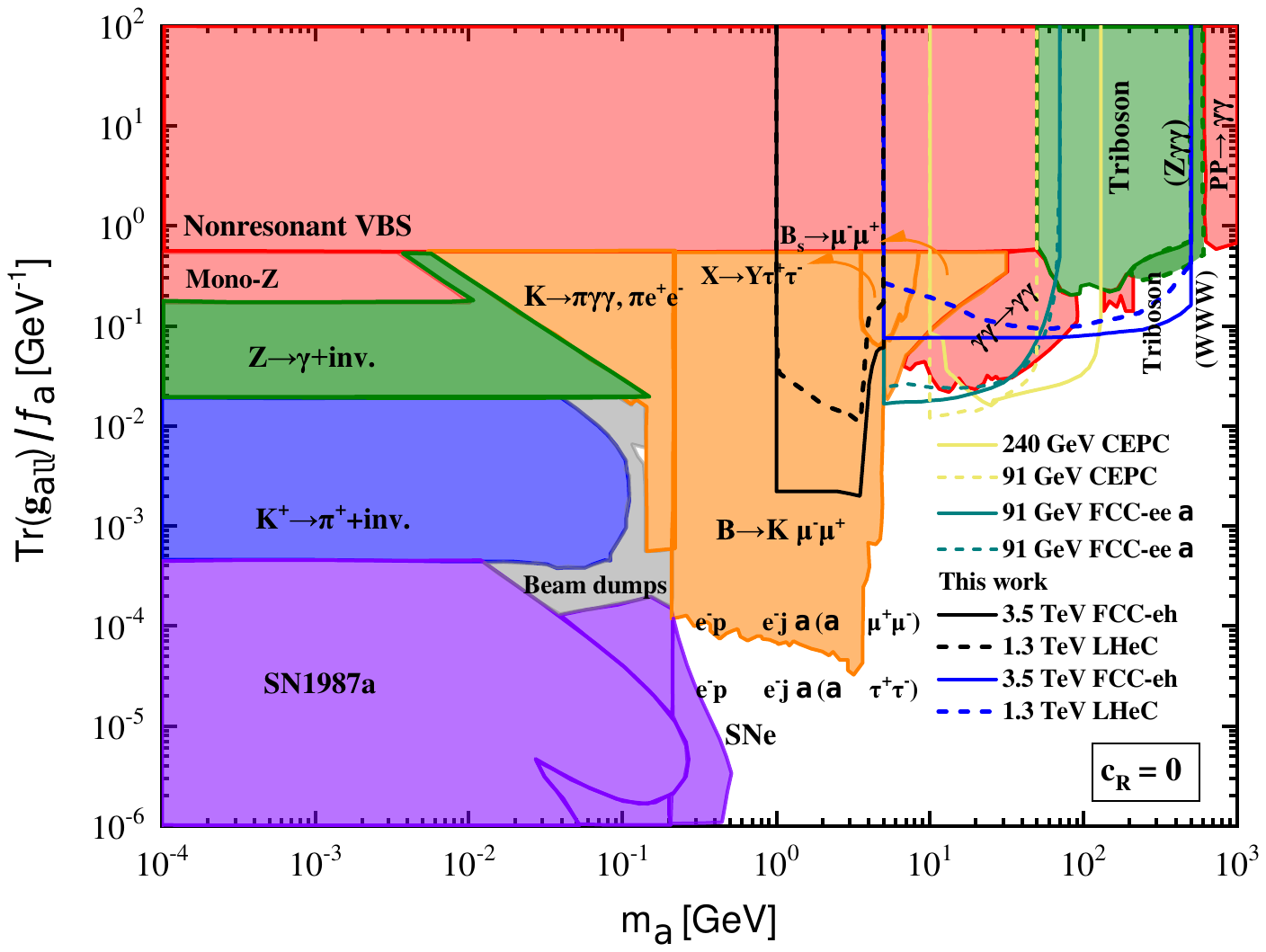}
\caption{For $c_{R}=0$, our projected $2\sigma$ sensitivities to the ALP-lepton coupling $Tr(g_{a\ell\ell})/f_a$ via the processes $e^{-}p\rightarrow e^-ja~(a \to \mu^+ \mu^-)$~(black dashed and solid lines) and $e^{-}p\rightarrow e^-ja~(a \to \tau^+ \tau^-, \tau^+ \tau^- \to \ell^+ \ell^- \slashed E, \ell = e,\mu)$~(blue dashed and solid lines) at the LHeC~(FCC-eh) with $\sqrt{s}=1.3~(3.5)$ TeV and $\mathcal{L}=$ $1~(2)$ ab$^{-1}$ and other current or expected excluded regions.}
\label{fig:13}
\end{center}
\end{figure}

The projected sensitivities of the $1.3$~($3.5$) TeV LHeC~(FCC-eh) with $\mathcal{L}=$ $1~(2)$ ab$^{-1}$ to the ALP-lepton coupling at 2$\sigma$ CL via the processes $e^{-}p\rightarrow e^-ja~(a \to \mu^+ \mu^-)$~(black dashed and solid lines) and $e^{-}p\rightarrow e^-ja~(a \to \tau^+ \tau^-, \tau^+ \tau^- \to \ell^+ \ell^- \slashed E, \ell = e,\mu)$~(blue dashed and solid lines) obtained in \ref{subsec:first} and \ref{subsec:second} are given in Fig.~\ref{fig:13}, where the current and expected exclusion regions for this coupling are also shown. The purple areas indicate the constraints from supernova observations, which are labeled as ``SNe" and ``SN1987a"~\cite{Ferreira:2022xlw,Diamond:2023scc,Caputo:2022mah}. The gray area depicts the constraints from beam dump experiments~\cite{CHARM:1985anb,Riordan:1987aw,Blumlein:1990ay,Dolan:2017osp,NA64:2020qwq,Waites:2022tov}.
The blue and orange areas respectively depict the results from searches for ALPs via rare decays of the mesons $K$ and $B$~\cite{NA62:2021zjw,E949:2005qiy,NA62:2014ybm,NA48:2002xke,KTeV:2008nqz,LHCb:2015nkv,LHCb:2016awg,Albrecht:2019zul,BaBar:2012sau}. The red and green areas represent the constraints from various searches at collider experiments, including the process $Z \to \gamma + \text{inv.}$ observed at the LEP (labeled ``$Z \to \gamma + \text{inv.}$")~\cite{Craig:2018kne,L3:1997exg}, mono-$Z$ event searches at the LHC (labeled ``mono-$Z$")~\cite{Brivio:2017ije} and nonresonant ALPs searches via vector boson scattering (VBS) processes at the LHC (labeled ``nonresonant VBS")~\cite{Bonilla:2022pxu}, light-by-light scattering ($\gamma\gamma \to \gamma\gamma$) in Pb-Pb collisions at the LHC~(labeled ``$\gamma\gamma \to \gamma\gamma$")~\cite{CMS:2018erd,ATLAS:2020hii}, diphoton production at the LHC~(labeled ``$p p \to \gamma\gamma$")~\cite{Mariotti:2017vtv} as well as triboson searches at the LHC~(labeled ``triboson $(WWW)$" and ``triboson $(Z\gamma\gamma)$")~\cite{CMS:2019mpq,Craig:2018kne}. The expected exclusion regions for the ALP-lepton couplings are derived from the processes $Z\to \mu^{+} \mu^{-} \slashed E$ and $Z\to e^+ e^- \mu^{+} \mu^{-}$ at the FCC-ee~\cite{Yue:2022ash}~(dark cyan solid and dashed lines) and the process $e^+ e^- \rightarrow \gamma\gamma E \mkern-10.5 mu/$ at the $240$~(91) GeV CEPC~\cite{Yue:2024xrc}~(yellow solid and dashed lines).

Our results show that the projected sensitivities obtained at the $1.3~(3.5)$ TeV LHeC~(FCC-eh) are in the ranges of $0.011~(0.002)$ $\sim$ $0.172~(0.060)$ GeV$^{-1}$ for $1~\text{GeV} \leq m_a \leq 5~\text{GeV}$ via the process $e^{-}p \rightarrow e^- j a \, (a \to \mu^+ \mu^-)$ and $0.094~(0.075)$ $\sim$ $0.440~(0.160)$ GeV$^{-1}$ for $5~\text{GeV} \leq m_a \leq 500~\text{GeV}$ via the process $e^- p \to e^- j a$~($a \to \tau^+ \tau^-$, $\tau^+ \tau^- \to \ell^+ \ell^- \slashed E$). In Table~\ref{tab:ALP_sensitivities}, we summarize projected sensitivities derived in this work to the ALP-lepton coupling $Tr(g_{a\ell\ell})/f_a$ and other results. Through comparison, we can draw a conclusion that the projected sensitivities obtained at the $1.3~(3.5)$ TeV LHeC~(FCC-eh) are stronger than the bounds from nonresonant ALP searches via VBS processes at the LHC, but weaker than the constraints from the rare meson decay process $B \to K \mu^- \mu^+$ for the ALPs with masses in the range of $1$ $\sim$ $5$ GeV. For the ALPs with masses in the range of $5$ $\sim$ $80~(70)$ GeV, the projected sensitivities obtained at the $1.3~(3.5)$ TeV LHeC~(FCC-eh) are excluded by the bounds from light-by-light scattering at the LHC. However, for $m_a$ in the range of $80~(70)$ $\sim$ $500$ GeV, the projected sensitivities obtained at the $1.3~(3.5)$ TeV LHeC~(FCC-eh) further complement the results of light-by-light scattering at the LHC, offering stronger constraints than those derived from triboson final state searches and diphoton production at the LHC. Compared to the results given by Refs.~\cite{Yue:2022ash,Yue:2024xrc} at future $e^+ e^-$ colliders, our signal process at the $1.3~(3.5)$ TeV LHeC~(FCC-eh) can explore a broader range of ALP masses, from $120~(110)$ to 500 GeV.

\begin{table}[H]\scriptsize
\begin{center}
\caption{For $c_{R}=0$, our projected $2\sigma$ sensitivities to the ALP-lepton coupling $Tr(g_{a\ell\ell})/f_a$ at the LHeC~(FCC-eh) and other results.}
\label{tab:ALP_sensitivities}
\resizebox{14cm}{!}{
\setlength{\tabcolsep}{10pt}
\begin{tabular}{|c|c|c|}
\hline
Experiment        & ~~ALP mass range~(GeV)~~       & ~~limits or expected limits (GeV$^{-1}$)~~ \\ \hline
Rare meson decay~($B \to K \mu^- \mu^+$)~\cite{LHCb:2015nkv,LHCb:2016awg} &  $0.2 \sim 4.7$ & $0.00003 \sim 0.0001$ \\ \hline
Rare meson decays~($B \to K \tau^+ \tau^-$ and $\Upsilon \to \gamma \tau^+ \tau^-$)~\cite{BaBar:2012sau} &  $3.6 \sim 9.2$ & $ > 0.090 $ \\ \hline
Non-resonant ALP searches via VBS processes at the LHC~\cite{Bonilla:2022pxu} & $0.0001 \sim 100$ & $> 0.560$ \\ \hline
Light-by-light scattering at the LHC~\cite{CMS:2018erd,ATLAS:2020hii} & $5 \sim 90$ & $0.022 \sim 0.132$ \\ \hline
Triboson $Z\gamma\gamma$ final state searches at the LHC~\cite{Craig:2018kne} & $20 \sim 500$ & $0.209 \sim 0.704$ \\ \hline
Triboson $WWW$ final state searches at the LHC~\cite{CMS:2019mpq} & $200 \sim 600$ & $0.360 \sim 0.513$ \\ \hline
Diphoton production at the LHC~\cite{Mariotti:2017vtv,Bonilla:2023dtf} & $130 \sim 1000$ & $0.140 \sim 0.677$ \\ \hline
$e^+ e^- \rightarrow \gamma\gamma E \mkern-10.5 mu/$ at the 91 GeV CEPC~\cite{Yue:2024xrc} &  $10 \sim 50$ & $0.012 \sim 0.044$ \\ \hline
$e^+ e^- \rightarrow \gamma\gamma E \mkern-10.5 mu/$ at the 240 GeV CEPC~\cite{Yue:2024xrc} & $10 \sim 130$ & $0.016 \sim 0.124$ \\ \hline
$Z\to \mu^{+} \mu^{-} \slashed E$ at the FCC-ee~\cite{Yue:2022ash} & $5 \sim 70$ & $0.017\sim 0.481$ \\ \hline
$Z\to e^+ e^- \mu^{+} \mu^{-}$ at the FCC-ee~\cite{Yue:2022ash} & $5 \sim 70$ & $0.024 \sim 3.608$ \\ \hline

$e^{-}p\rightarrow e^-ja~(a \to \mu^+ \mu^-)$ at the LHeC & $1 \sim 5$ & $0.011 \sim 0.172$ \\ \hline

$e^{-}p\rightarrow e^-ja~(a \to \mu^+ \mu^-)$ at the FCC-eh & $1 \sim 5$ & $0.002 \sim 0.060$ \\ \hline

$e^{-}p\rightarrow e^-ja~(a \to \tau^+ \tau^-)$ at the LHeC & $5 \sim 500$  & $0.094 \sim 0.440$ \\ \hline

$e^{-}p\rightarrow e^-ja~(a \to \tau^+ \tau^-)$ at the FCC-eh & $5 \sim 500$ & $0.075 \sim 0.160$ \\ \hline
\end{tabular}}
\end{center}
\end{table}

\subsection{Searching for the ALP-lepton couplings via three-lepton final state processes for $c_{L}=0$}

\subsubsection{The process $e^{-}p\rightarrow e^-ja~(a\rightarrow\mu^{+}\mu^{-})$}\label{subsec:third}

In this subsection, we also investigate the signal process $e^{-}p\rightarrow e^-ja(a\rightarrow\mu^{+}\mu^{-})$ but for $c_{L}=0$ (i.e. $Tr(g_{a\nu\nu})/f_a=0$) with efforts to obtain projected sensitivity to the ALP-lepton coupling $Tr(g_{a\ell\ell})/f_a$. The production cross sections of this process, varying with $m_a$ and $Tr(g_{a\ell\ell})/f_a$ at the $1.3~(3.5)$ TeV LHeC~(FCC-eh), follow a similar trend to Fig.\ref{fig:5} but show greater values. For $1$ GeV $ \leq m_a\leq5$ GeV, $Tr(g_{a\ell\ell})/f_a=0.1$ and $0.01$ GeV$^{-1}$, the production cross sections span from $3.690 \times 10^{-3}~(1.394 \times 10^{-2})$ to $1.840 \times 10^{-5}~(6.948 \times 10^{-5})$ pb and
$3.703 \times 10^{-5}~(1.392 \times 10^{-4})$ to $1.862 \times 10^{-7}~(6.973 \times 10^{-7})$ pb, respectively. These numerical results have employed the basic cuts mentioned above.
The main SM background has been discussed in \ref{subsec:first}. The kinematic distributions for the signal and background at the $1.3$~($3.5$) TeV LHeC~(FCC-eh) with $\mathcal{L}=$ $1$~($2$) ab$^{-1}$ are similar to those for $c_{R}=0$, which are not shown in this subsection. So we employ the same improved cut in this case. In Tables~\ref{tab7} and~\ref{tab8}, we summarized the cross sections of the signal and background after imposing improved cuts for few ALP mass benchmark points and the parameter $Tr(g_{a\ell\ell})/f_a=0.1$ GeV$^{-1}$ at the $1.3~(3.5)$ TeV LHeC~(FCC-eh) with $\mathcal{L}=$ $1~(2)$ ab$^{-1}$. The corresponding $SS$ values of benchmark points are also shown in these tables. The $SS$ values can range from $2.785~(159.229)$ to $0.830~(8.427)$ for $m_a=1$ $\sim$ $5$ GeV.

\begin{table}[H]\tiny
	\centering{
\caption{The production cross sections of the signal and SM background after the improved cuts applied at the $1.3$ TeV LHeC with $\mathcal{L}=$ $1$ ab$^{-1}$ for $Tr(g_{a\ell\ell})/f_a=0.1$ GeV$^{-1}$ with the benchmark points $m_a = 1$, $2$, $3$, $4$, $5$ GeV.$~$\label{tab7}}
		\newcolumntype{C}[1]{>{\centering\let\newline\\\arraybackslash\hspace{50pt}}m{#1}}
		\begin{tabular}{m{1.5cm}<{\centering}|m{2.3cm}<{\centering} m{2.3cm}<{\centering} m{2.3cm}<{\centering}  m{2.3cm}<{\centering} m{2.3cm}<{\centering}}
			\hline \hline
      \multirow{2}{*}{Cuts} & \multicolumn{5}{c}{cross sections for signal (background) [pb]}\\
     \cline{2-6}
     & $m_a=1$ GeV  & $m_a=2$ GeV  & $m_a=3$ GeV  & $m_a=4$ GeV & $m_a=5$ GeV  \\ \hline
     Basic Cuts  & \makecell{$3.709\times10^{-3}$\\$(0.041)$} & \makecell{$3.704\times10^{-3}$\\$(0.041)$} &\makecell{$3.697\times10^{-3}$\\$(0.041)$} &\makecell{$2.819\times10^{-5}$\\$(0.041)$} & \makecell{$1.841\times10^{-5}$\\$(0.041)$}
        \\
     Cut 1  & \makecell{$9.860\times10^{-6}$\\$(0.019)$} & \makecell{$1.063\times10^{-4}$\\$(0.019)$} &\makecell{$2.900\times10^{-4}$\\$(0.019)$} &\makecell{$3.990\times10^{-6}$\\$(0.019)$} & \makecell{$3.850\times10^{-6}$\\$(0.019)$}
       \\
     Cut 2  & \makecell{$8.380\times10^{-6}$\\$(3.373\times10^{-3})$} & \makecell{$1.003\times10^{-4}$\\$(3.373\times10^{-3})$} &\makecell{$2.660\times10^{-4}$\\$(3.373\times10^{-3})$} &\makecell{$3.570\times10^{-6}$\\$(3.373\times10^{-3})$} & \makecell{$3.380\times10^{-6}$\\$(3.373\times10^{-3})$}
       \\
     Cut 3  & \makecell{$8.140\times10^{-6}$\\$(4.090\times10^{-7})$} & \makecell{$9.976\times10^{-5}$\\$(1.640\times10^{-7})$} &\makecell{$2.627\times10^{-4}$\\$(2.460\times10^{-7})$} &\makecell{$3.460\times10^{-6}$\\$(8.190\times10^{-7})$} & \makecell{$3.180\times10^{-6}$\\$(1.146\times10^{-5})$}
     \\ \hline
     $SS$  & $2.785$ & $9.980$ & $16.202$ & $1.672$ & $0.830$ \\ \hline \hline
	\end{tabular}}	
\end{table}

\begin{table}[H]\tiny
	\centering{
\caption{Same as Table~\ref{tab7} but for the  $3.5$ TeV FCC-eh with $\mathcal{L}=$ $2$ ab$^{-1}$.$~$\label{tab8}}
		\newcolumntype{C}[1]{>{\centering\let\newline\\\arraybackslash\hspace{50pt}}m{#1}}
		\begin{tabular}{m{1.5cm}<{\centering}|m{2.3cm}<{\centering} m{2.3cm}<{\centering} m{2.3cm}<{\centering}  m{2.3cm}<{\centering} m{2.3cm}<{\centering}}
			\hline \hline
      \multirow{2}{*}{Cuts} & \multicolumn{5}{c}{cross sections for signal (background) [pb]}\\
     \cline{2-6}
     & $m_a=1$ GeV  & $m_a=2$ GeV  & $m_a=3$ GeV  & $m_a=4$ GeV & $m_a=5$ GeV  \\ \hline
     Basic Cuts  & \makecell{$1.398\times10^{-2}$\\$(0.095)$} & \makecell{$1.397\times10^{-2}$\\$(0.095)$} &\makecell{$1.396\times10^{-2}$\\$(0.095)$} &\makecell{$1.065\times10^{-4}$\\$(0.095)$} & \makecell{$6.959\times10^{-5}$\\$(0.095)$}
        \\
     Cut 1  & \makecell{$1.270\times10^{-2}$\\$(0.045)$} & \makecell{$1.240\times10^{-2}$\\$(0.045)$} &\makecell{$1.220\times10^{-2}$\\$(0.045)$} &\makecell{$9.303\times10^{-5}$\\$(0.045)$} & \makecell{$6.058\times10^{-5}$\\$(0.045)$}
       \\
     Cut 2  & \makecell{$1.270\times10^{-2}$\\$(2.457\times10^{-3})$} & \makecell{$1.240\times10^{-2}$\\$(2.457\times10^{-3})$} &\makecell{$1.210\times10^{-2}$\\$(2.457\times10^{-3})$} &\makecell{$8.980\times10^{-5}$\\$(2.457\times10^{-3})$} & \makecell{$5.680\times10^{-5}$\\$(2.457\times10^{-3})$}
       \\
     Cut 3  & \makecell{$1.260\times10^{-2}$\\$(1.140\times10^{-6})$} & \makecell{$1.230\times10^{-2}$\\$(9.400\times10^{-7})$} &\makecell{$1.200\times10^{-2}$\\$(7.600\times10^{-7})$} &\makecell{$8.916\times10^{-5}$\\$(7.600\times10^{-7})$} & \makecell{$5.619\times10^{-5}$\\$(3.273\times10^{-5})$}
     \\ \hline
     $SS$  & $159.229$ & $157.056$ & $154.997$ & $13.297$ & $8.427$ \\ \hline \hline
	\end{tabular}}	
\end{table}

\begin{figure}[H]
\begin{center}
\centering\includegraphics [scale=0.33] {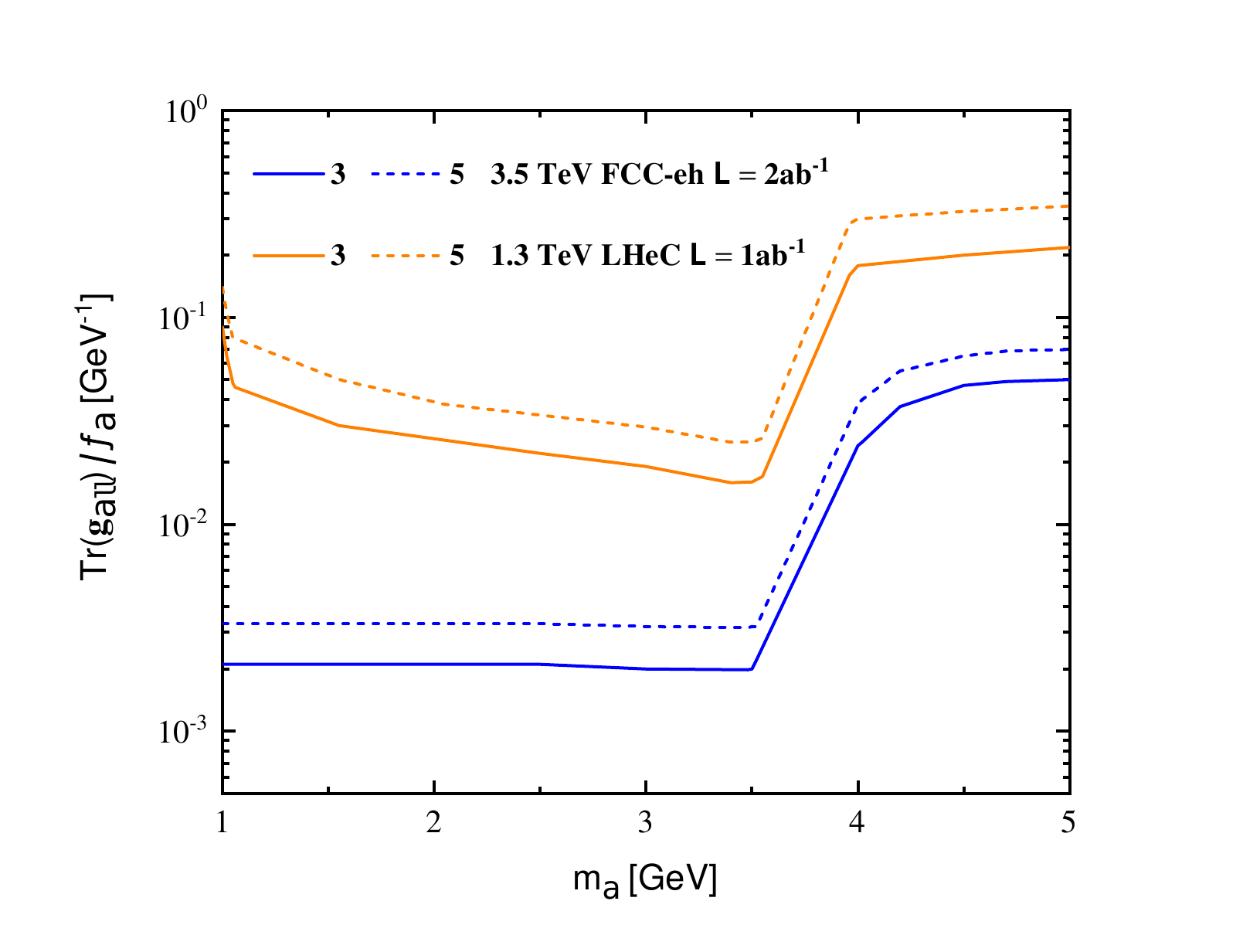}
\caption{The $3\sigma$ and $5\sigma$ curves in the $m_a - Tr(g_{a\ell\ell})/f_a$ plane for the three-lepton final state process $e^- p \to e^- j a~(a \to \mu^+ \mu^-)$ at the $1.3$~($3.5$) TeV LHeC~(FCC-eh) with $\mathcal{L}=$ $1$~($2$) ab$^{-1}$ for $c_{L}=0$.}
\label{fig:14}
\end{center}
\end{figure}

In Fig.~\ref{fig:14}, we plot the $3\sigma$ and $5\sigma$ curves in the plane of $m_a - Tr(g_{a\ell\ell})/f_a$ for the $1.3$ TeV LHeC with $\mathcal{L}=$ $1$ ab$^{-1}$~(orange solid and dashed lines) and $3.5$ TeV FCC-eh with $\mathcal{L}=$ $2$ ab$^{-1}$~(blue solid and dashed lines). From Fig.~\ref{fig:14}, we can see that the prospective sensitivities respectively as $0.016$~($0.002$)  GeV$^{-1}$ $\leq$ $Tr(g_{a\ell\ell})/f_a$ $\leq$ $0.218$~($0.050$) GeV$^{-1}$ and $0.025$~($0.003$) GeV$^{-1}$ $\leq$ $Tr(g_{a\ell\ell})/f_a$ $\leq$ $0.345$~($0.070$) GeV$^{-1}$ at 3$\sigma$ and 5$\sigma$ CL for the ALPs with masses in the range of $1$ $\sim$ $5$ GeV at the $1.3$~($3.5$) TeV LHeC~(FCC-eh) with $\mathcal{L}=$ $1$~($2$) ab$^{-1}$.
The $1.3$~($3.5$) TeV LHeC~(FCC-eh) can provide better prospective sensitivities to the ALP-lepton couplings via the process $e^{-}p\rightarrow e^-ja~(a\rightarrow\mu^{+}\mu^{-})$ for $c_L = 0$ compared to those for $c_R = 0$.

\subsubsection{The process $e^{-}p\rightarrow e^-ja~(a \to \tau^+ \tau^-, \tau^+ \tau^- \to \ell^+ \ell^{-} \slashed E, \ell= e,\mu)$}\label{subsec:forth}

In this subsection, we will consider the same signal process as described in \ref{subsec:second} with the goal of obtaining prospective sensitivities to the ALP-lepton coupling $Tr(g_{a\ell\ell})/f_a$ for $c_{L}=0$. The tendency of the production cross sections for the signal process as varying with the ALP mass and the ALP-lepton coupling at the $1.3$~(3.5) TeV LHeC~(FCC-eh) is same as the case for $c_{R}=0$, which is not presented again in this paper. For $5$ GeV $ \leq m_a\leq 500$ GeV, $Tr(g_{a\ell\ell})/f_a=0.1$ and $0.01$ GeV$^{-1}$, the production cross sections of the signal process can range from $3.650 \times 10^{-3}~(1.384 \times 10^{-2})$ to $5.540 \times 10^{-5}~(1.660\times 10^{-3})$ pb and $3.658 \times 10^{-5}~(1.387 \times 10^{-4})$ to $5.520 \times 10^{-7}~(1.655\times 10^{-5})$ pb at the $1.3~(3.5)$ TeV LHeC~(FCC-eh), respectively. These numerical results have been obtained with the basic cuts, as outlined in the preceding sections. The corresponding SM background is same as that in the case of $c_{R}=0$.

The kinematic distributions for the signal and background are similar to those discussed earlier, so we apply the same improved cut in this case. In Tables~\ref{tab9} and~\ref{tab10}, we summarized the cross sections of the signal and background after imposing  improved cuts for few ALP mass benchmark points and the parameter $Tr(g_{a\ell\ell})/f_a=0.1$ GeV$^{-1}$. The corresponding $SS$ values of benchmark points are also shown in these tables. The $SS$ values can range from $2.874~(9.804)$ to $0.209~(3.751)$ for $m_a=5$ $\sim$ $500$ GeV at the $1.3~(3.5)$ TeV LHeC~(FCC-eh) with $\mathcal{L}=$ $1~(2)$ ab$^{-1}$.

\begin{table}[H]\tiny
	\centering{
\caption{The production cross sections of the signal and SM background after the improved cuts applied for $Tr(g_{a\ell\ell})/f_a=0.1$ GeV$^{-1}$ at the $1.3$ TeV LHeC with $\mathcal{L}=$ $1$ ab$^{-1}$ with benchmark points $m_a = 50$, $170$, $290$, $410$, $500$ GeV.$~$\label{tab9}}
		\newcolumntype{C}[1]{>{\centering\let\newline\\\arraybackslash\hspace{50pt}}m{#1}}
		\begin{tabular}{m{1.5cm}<{\centering}|m{2.3cm}<{\centering} m{2.3cm}<{\centering} m{2.3cm}<{\centering}  m{2.3cm}<{\centering} m{2.3cm}}
			\hline \hline
      \multirow{2}{*}{Cuts} & \multicolumn{5}{c}{cross sections for signal (background) [pb]}\\
     \cline{2-6}
     & $m_a=50$ GeV  & $m_a=170$ GeV  & $m_a=290$ GeV & $m_a=410$ GeV &  $m_a=500$ GeV   \\ \hline
     Basic Cuts  & \makecell{$3.133\times10^{-4}$\\$(0.011)$} & \makecell{$1.097\times10^{-4}$\\$(0.011)$} &\makecell{$4.083\times10^{-5}$\\$(0.011)$} &\makecell{$1.483\times10^{-5}$\\$(0.011)$} & \makecell{$6.582\times10^{-6}$\\$(0.011)$}
        \\
      Cut 1  & \makecell{$2.757\times10^{-4}$\\$(8.257\times10^{-3})$} & \makecell{$9.680\times10^{-5}$\\$(8.257\times10^{-3})$} &\makecell{$3.580\times10^{-5}$\\$(8.257\times10^{-3})$} &\makecell{$1.292\times10^{-5}$\\$(8.257\times10^{-3})$} & \makecell{$5.711\times10^{-6}$\\$(8.257\times10^{-3})$}
       \\
     Cut 2  &\makecell{$1.065\times10^{-4}$\\$(1.109\times10^{-3})$}  & \makecell{$5.604\times10^{-5}$\\$(1.109\times10^{-3})$} &\makecell{$2.876\times10^{-5}$\\$(1.109\times10^{-3})$} &\makecell{$1.146\times10^{-5}$\\$(1.109\times10^{-3})$} & \makecell{$5.240\times10^{-6}$\\$(1.109\times10^{-3})$}
       \\
     Cut 3  & \makecell{$7.146\times10^{-5}$\\$(5.836\times10^{-4})$} & \makecell{$4.543\times10^{-5}$\\$(5.836\times10^{-4})$} &\makecell{$2.582\times10^{-5}$\\$(5.836\times10^{-4})$} &\makecell{$1.074\times10^{-5}$\\$(5.836\times10^{-4})$} & \makecell{$5.000\times10^{-6}$\\$(5.836\times10^{-4})$}
     \\
     Cut 4  & \makecell{$6.363\times10^{-5}$\\$(4.264\times10^{-4})$} & \makecell{$3.860\times10^{-5}$\\$(4.264\times10^{-4})$} &\makecell{$2.185\times10^{-5}$\\$(4.264\times10^{-4})$} &\makecell{$9.250\times10^{-6}$\\$(4.264\times10^{-4})$} & \makecell{$4.330\times10^{-6}$\\$(4.264\times10^{-4})$}
     \\ \hline
     $SS$  & $2.874$ & $1.790$ & $1.032$ & $0.443$ & ~~~~~~~~~~~$0.209$  \\ \hline \hline
	\end{tabular}}	
\end{table}

\begin{table}[H]\tiny
	\centering{
\caption{Same as Table~\ref{tab9} but for the $3.5$ TeV FCC-eh with $\mathcal{L}=$ $2$ ab$^{-1}$.$~$\label{tab10}}
		\newcolumntype{C}[1]{>{\centering\let\newline\\\arraybackslash\hspace{50pt}}m{#1}}
		\begin{tabular}{m{1.5cm}<{\centering}|m{2.3cm}<{\centering} m{2.3cm}<{\centering} m{2.3cm}<{\centering}  m{2.3cm}<{\centering} m{2.3cm}}
			\hline \hline
      \multirow{2}{*}{Cuts} & \multicolumn{5}{c}{cross sections for signal (background) [pb]}\\
     \cline{2-6}
     & $m_a=50$ GeV  & $m_a=170$ GeV  & $m_a=290$ GeV & $m_a=410$ GeV &  $m_a=500$ GeV   \\ \hline
     Basic Cuts  & \makecell{$1.340\times10^{-3}$\\$(0.037)$} & \makecell{$7.249\times10^{-4}$\\$(0.037)$} &\makecell{$4.348\times10^{-4}$\\$(0.037)$} &\makecell{$2.744\times10^{-4}$\\$(0.037)$} & \makecell{$1.974\times10^{-4}$\\$(0.037)$}
        \\
     Cut 1  & \makecell{$1.178\times10^{-3}$\\$(0.029)$} & \makecell{$6.455\times10^{-4}$\\$(0.029)$} &\makecell{$3.855\times10^{-4}$\\$(0.029)$} &\makecell{$2.420\times10^{-4}$\\$(0.029)$} & \makecell{$1.736\times10^{-4}$\\$(0.029)$}
       \\
     Cut 2  &\makecell{$5.637\times10^{-4}$\\$(5.711\times10^{-3})$}  & \makecell{$3.938\times10^{-4}$\\$(5.711\times10^{-3})$} &\makecell{$3.168\times10^{-4}$\\$(5.711\times10^{-3})$} &\makecell{$2.201\times10^{-4}$\\$(5.711\times10^{-3})$} & \makecell{$1.631\times10^{-4}$\\$(5.711\times10^{-3})$}
       \\
     Cut 3  & \makecell{$4.320\times10^{-4}$\\$(3.765\times10^{-3})$} & \makecell{$3.380\times10^{-4}$\\$(3.765\times10^{-3})$} &\makecell{$2.934\times10^{-4}$\\$(3.765\times10^{-3})$} &\makecell{$2.102\times10^{-4}$\\$(3.765\times10^{-3})$} & \makecell{$1.578\times10^{-4}$\\$(3.765\times10^{-3})$}
     \\
     Cut 4  & \makecell{$3.919\times10^{-4}$\\$(2.803\times10^{-3})$} & \makecell{$3.045\times10^{-4}$\\$(2.803\times10^{-3})$} &\makecell{$2.607\times10^{-4}$\\$(2.803\times10^{-3})$} &\makecell{$1.897\times10^{-4}$\\$(2.803\times10^{-3})$} & \makecell{$1.440\times10^{-4}$\\$(2.803\times10^{-3})$}
     \\
     \hline
     $SS$  & $9.804$ & $7.725$ & $6.661$ & $4.904$ & ~~~~~~~~~~~$3.751$  \\ \hline \hline
	\end{tabular}}	
\end{table}

\begin{figure}[H]
\begin{center}
\centering\includegraphics [scale=0.3] {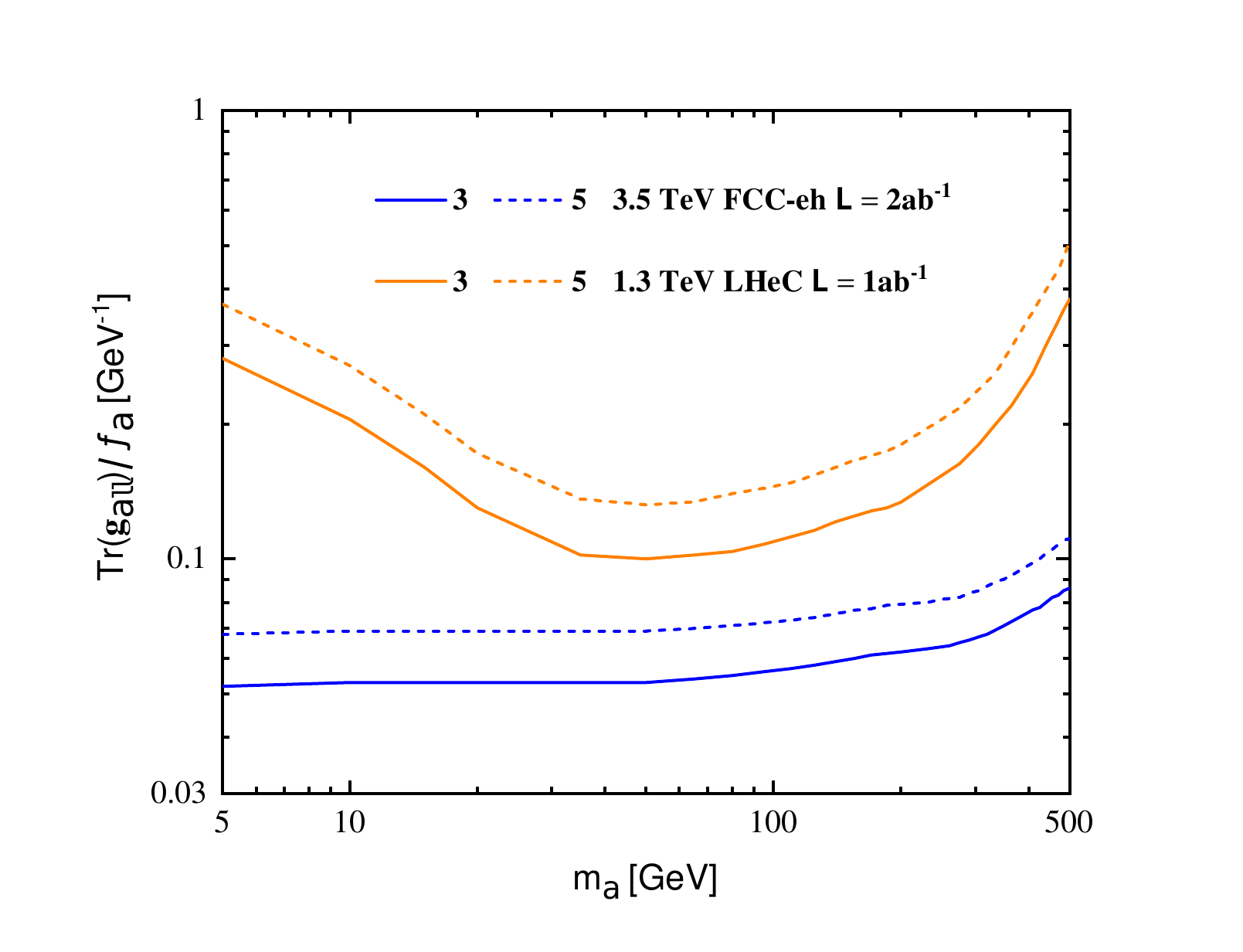}
\caption{The $3\sigma$ and $5\sigma$ curves in the $m_a - Tr(g_{a\ell\ell})/f_a$ plane for the three-lepton final state process $e^{-}p\rightarrow e^-ja~(a \to \tau^+ \tau^-, \tau^+ \tau^- \to \ell^+ \ell^{-} \slashed E, \ell= e,\mu)$ at the $1.3$~($3.5$) TeV LHeC~(FCC-eh) with $\mathcal{L}=$ $1$~($2$) ab$^{-1}$ for $c_{L}=0$.}
\label{fig:15}
\end{center}
\end{figure}

In FIG.~\ref{fig:15}, we plot the $3\sigma$ and $5\sigma$ curves in the plane of $m_a - Tr(g_{a\ell\ell})/f_a$ for the LHeC with $\sqrt{s}=1.3$ TeV and $\mathcal{L}=$ $1$ ab$^{-1}$~(orange solid and dashed lines) and FCC-eh with $\sqrt{s}=3.5$ TeV and $\mathcal{L}=$ $2$ ab$^{-1}$~(blue solid and dashed lines). As shown in Fig.~\ref{fig:15}, we can obtain that the prospective sensitivities respectively as $0.112$~$(0.052)$ GeV$^{-1}$ $\leq$ $Tr(g_{a\ell\ell})/f_a$ $\leq$ $0.380$~$(0.086)$ GeV$^{-1}$ and $0.132$~$(0.068)$ GeV$^{-1}$ $\leq$ $Tr(g_{a\ell\ell})/f_a$ $\leq$ $0.510$~$(0.111)$ GeV$^{-1}$ at $3$$\sigma$ and $5$$\sigma$ CL for the ALPs with masses in the range of $5$ $\sim$ $500$ GeV. Comparing to the calculated results derived in \ref{subsec:third}, it is evident that the LHeC~(FCC-eh) can obtain stronger expected limits on the ALP-lepton couplings for the ALPs with masses in the range of $1$ $\sim$ $5$ GeV via the process $e^{-}p \rightarrow e^- j a \, (a \to \mu^+ \mu^-)$ than those from the process $e^{-}p \rightarrow e^- j a \, (a \to \tau^+ \tau^-, \tau^+ \tau^- \to \ell^+ \ell^- \slashed{E}, \ell = e,\mu)$ for the ALPs with masses in the range of $5$ $\sim$ $500$ GeV. In addition, we can find that the $1.3$~($3.5$) TeV LHeC~(FCC-eh) can provide better prospective sensitivities to the ALP-lepton couplings via the process $e^{-}p \rightarrow e^- j a \, (a \to \tau^+ \tau^-, \tau^+ \tau^- \to \ell^+ \ell^- \slashed{E}, \ell = e,\mu)$ for $c_L = 0$ than those for $c_R = 0$.

\begin{figure}[H]
\begin{center}
\centering\includegraphics [scale=0.4] {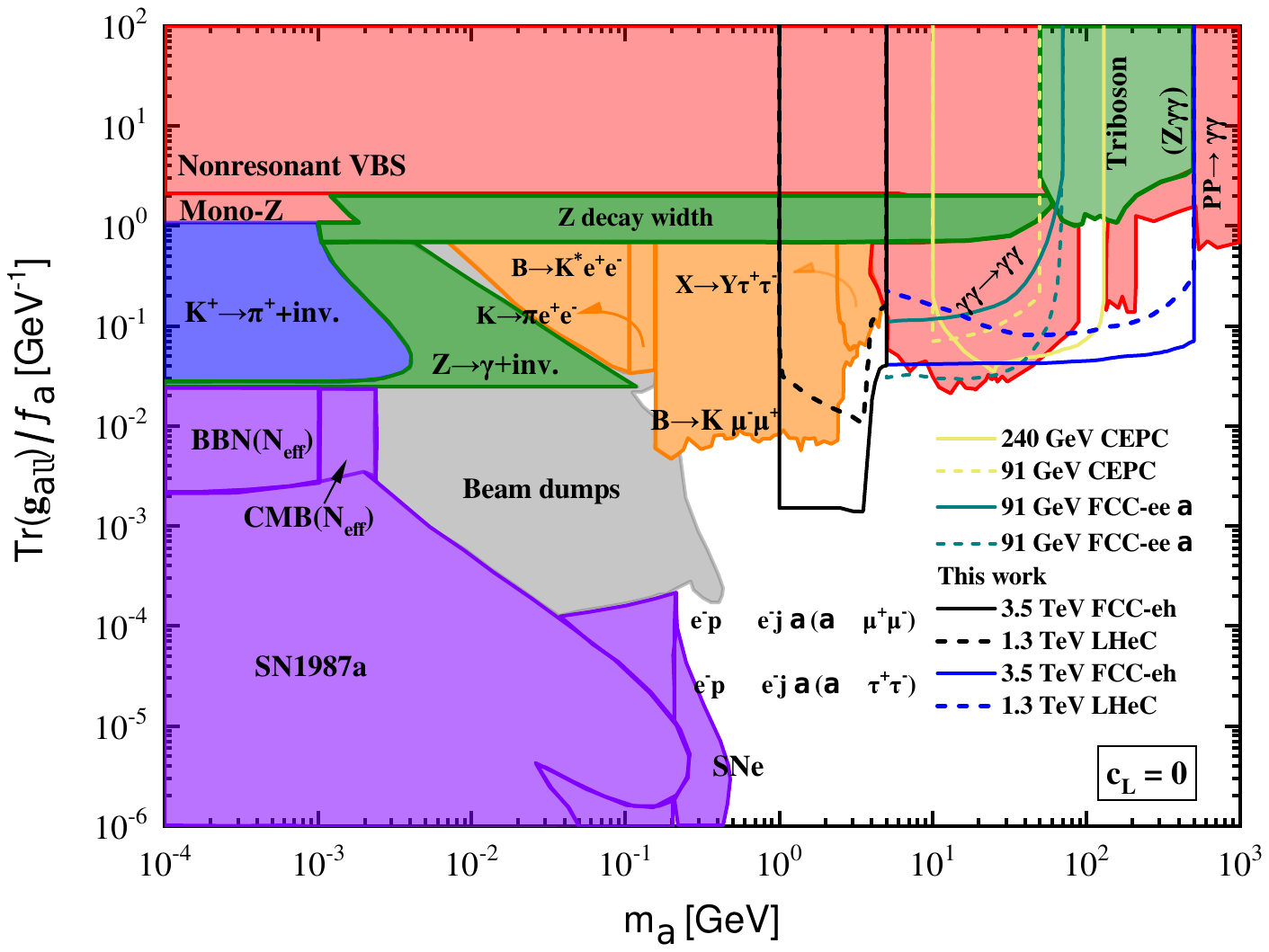}
\caption{For $c_{L}=0$, our projected $2\sigma$ sensitivities to the ALP-lepton coupling $Tr(g_{a\ell\ell})/f_a$ via the processes $e^{-}p\rightarrow e^-ja~(a \to \mu^+ \mu^-)$~(black dashed and solid lines) and $e^{-}p\rightarrow e^-ja~(a \to \tau^+ \tau^-, \tau^+ \tau^- \to \ell^+ \ell^{-} \slashed E, \ell= e,\mu)$~(blue dashed and solid lines) at the LHeC~(FCC-eh) with $\sqrt{s}=1.3~(3.5)$ TeV and $\mathcal{L}=$ $1~(2)$ ab$^{-1}$  and other current or expected excluded regions.}
\label{fig:16}
\end{center}
\end{figure}

The projected sensitivities of the $1.3$~($3.5$) TeV LHeC~(FCC-eh) with $\mathcal{L}=$ $1~(2)$ ab$^{-1}$ to the ALP-lepton coupling at $2\sigma$ CL via the processes $e^{-}p\rightarrow e^-ja~(a \to \mu^+ \mu^-)$~(black dashed and solid lines) and $e^{-}p\rightarrow e^-ja~(a \to \tau^+ \tau^-, \tau^+ \tau^- \to \ell^+ \ell^{-} \slashed E, \ell= e,\mu)$~(blue dashed and solid lines) obtained in \ref{subsec:third} and \ref{subsec:forth} are given in Fig.~\ref{fig:16}, where the current and expected exclusion regions for this coupling are also shown. The green area depicts the limits from measurements of the total $Z$ decay width at the LEP, which is labeled as ``$Z$ decay width"~\cite{Craig:2018kne,Brivio:2017ije}. The orange area gives the exclusion region from the decay process $B \to K^{*} e^{+} e^{-}$, which is labeled as ``$B \to K^{*} e^{+} e^{-}$"~\cite{LHCb:2015ycz}. The purple areas depict the bounds from CMB and BBN observations, which are labeled as ``CMB~($N_{eff}$)" and ``BBN~($N_{eff}$)"~\cite{Ghosh:2020vti,Depta:2020zbh}. In addition, the constraints from other experiments have already been discribed in the previous subsection.

\vspace{-10pt}

It can be seen that the projected sensitivities obtained at the $1.3~(3.5)$ TeV LHeC~(FCC-eh) are in the ranges of $0.011~(0.001)$ $\sim$ $0.164~(0.040)$ GeV$^{-1}$ for $1~\text{GeV} \leq m_a \leq 5~\text{GeV}$ via the process $e^{-}p \rightarrow e^- j a \, (a \to \mu^+ \mu^-)$ and $0.080~(0.041)$ $\sim$ $0.310~(0.070)$ GeV$^{-1}$ for $5~\text{GeV} \leq m_a \leq 500~\text{GeV}$ via the process $e^- p \to e^- j a$~($a \to \tau^+ \tau^-$, $\tau^+ \tau^- \to \ell^+ \ell^- \slashed E$). In Table~\ref{tab:ALP_sensitivities1}, we present a comparison of our projected sensitivities to the ALP-lepton coupling $Tr(g_{a\ell\ell})/f_a$ at the LHeC~(FCC-eh) with other results. We can draw a conclusion that the projected sensitivities obtained at the $1.3~(3.5)$ TeV LHeC~(FCC-eh) are stronger than the bounds from rare meson decays, LEP measurements of the total $Z$ decay width or nonresonant ALP searches via VBS processes at the LHC for the ALPs with masses in the range of $2.5~(1)$ $\sim$ $4~(5)$ GeV. Notably, compared to the constraints derived from the rare meson decay process $B \to K \mu^- \mu^+$, the projected sensitivities of the $3.5$ TeV FCC-eh are enhanced by approximately up to an order of magnitude.
For $m_a$ in the range of $80~(50)$ $\sim$ $500$ GeV, the projected sensitivities obtained at this work further complement the results of light-by-light scattering at the LHC, providing stronger constraints than those from nonresonant ALP searches via VBS processes, triboson final state searches, diphoton production at the LHC and LEP measurements of the total $Z$ decay width. Compared to the results given by Refs.~\cite{Yue:2022ash,Yue:2024xrc} at future $e^+ e^-$ colliders, our signal process at the $1.3~(3.5)$ TeV LHeC~(FCC-eh) can detect a broader range of ALP masses, from $120~(45)$ to $500$ GeV.

\begin{table}[H]
\begin{center}\scriptsize
\caption{For $c_{L}=0$, our projected $2\sigma$ sensitivities to the ALP-lepton coupling $Tr(g_{a\ell\ell})/f_a$ at the LHeC~(FCC-eh) and other results.}
\label{tab:ALP_sensitivities1}
\resizebox{14cm}{!}{
\setlength{\tabcolsep}{10pt}
\begin{tabular}{|c|c|c|}
\hline
Experiment        & ~~ALP mass range~(GeV)~~       & ~~limits or expected limits (GeV$^{-1}$)~~ \\ \hline
Rare meson decay~($B \to K \mu^- \mu^+$)~\cite{LHCb:2015nkv,LHCb:2016awg} &  $0.2 \sim 4.7$ & $0.005 \sim 0.063$ \\ \hline
Rare meson decays~($B \to K \tau^+ \tau^-$ and $\Upsilon \to \gamma \tau^+ \tau^-$)~\cite{BaBar:2012sau} &  $3.6 \sim  9.2$ & $ > 0.058$ \\ \hline
Non-resonant ALP searches via VBS processes at the LHC~\cite{Bonilla:2022pxu} & $ 0.0001 \sim 100$ & $> 1.993$ \\ \hline
Measurements of the total $Z$ decay width at the LEP~\cite{Craig:2018kne,Brivio:2017ije} & $ 0.001 \sim 55$ & $0.687 \sim 1.665$ \\ \hline
Light-by-light scattering at the LHC~\cite{CMS:2018erd,ATLAS:2020hii} & $5 \sim 90$ & $0.021 \sim 0.111$ \\ \hline
Triboson $Z\gamma\gamma$ final state searches at the LHC~\cite{Craig:2018kne} & $20 \sim 500$ & $0.990 \sim 3.666$ \\ \hline
Diphoton production at the LHC~\cite{Mariotti:2017vtv,Bonilla:2023dtf} & $130 \sim 1000$ & $0.142 \sim 1.581$ \\ \hline
$e^+ e^- \rightarrow \gamma\gamma E \mkern-10.5 mu/$ at the 91 GeV CEPC~\cite{Yue:2024xrc} &  $10 \sim 50$ & $0.070 \sim 0.220$ \\ \hline
$e^+ e^- \rightarrow \gamma\gamma E \mkern-10.5 mu/$ at the 240 GeV CEPC~\cite{Yue:2024xrc} & $10 \sim 130$ & $0.035 \sim 0.180$ \\ \hline
$Z\to \mu^{+} \mu^{-} \slashed E$ at the FCC-ee~\cite{Yue:2022ash} & $5 \sim 70$ & $0.109\sim 3.161$ \\ \hline
$Z\to e^+ e^- \mu^{+} \mu^{-}$ at the FCC-ee~\cite{Yue:2022ash} & $5 \sim 70$ & $0.030 \sim 4.480$ \\ \hline

$e^{-}p\rightarrow e^-ja~(a \to \mu^+ \mu^-)$ at the LHeC & $1 \sim 5$ & $0.011 \sim 0.164$ \\ \hline

$e^{-}p\rightarrow e^-ja~(a \to \mu^+ \mu^-)$ at the FCC-eh & $1 \sim 5$ & $0.001 \sim 0.040$ \\ \hline

$e^{-}p\rightarrow e^-ja~(a \to \tau^+ \tau^-)$ at the LHeC & $5 \sim 500$  & $0.080 \sim 0.310$ \\ \hline

$e^{-}p\rightarrow e^-ja~(a \to \tau^+ \tau^-)$ at the FCC-eh & $5 \sim 500$ & $0.041 \sim 0.070$ \\ \hline
\end{tabular}}
\end{center}
\end{table}

\section{Summary and conclusion}

In recent years, extensive studies have focused on searching for new physics beyond the standard model. ALP is a promising candidate with strong theoretical motivation and may  be  the first new particle to be discovered at high-energy collider experiments. Future electron-proton colliders, with high center-of-mass energy and clean environment, are capable of probing a wider range of accessible parameter space for the ALP-lepton couplings.

In this paper, we consider the possibility of detecting the ALP-lepton couplings through three-lepton final state processes $e^- p \to e^- j a~(a \to \mu^+ \mu^-)$ and $e^{-}p\rightarrow e^-ja~(a \to \tau^+ \tau^-, \tau^+ \tau^- \to \ell^+ \ell^{-} \slashed E, \ell= e,\mu)$ for the ALPs with masses in the ranges of $1$ $\sim$ $5$ GeV and $5$ $\sim$ $500$ GeV at the $1.3~(3.5)$ TeV LHeC (FCC-eh) with $\mathcal{L}=$ $1~(2)$ ab$^{-1}$ for $c_{R}=0$ and $c_{L}=0$. We have performed simulations and analyses for the signal and relevant SM background processes and obtained $2$$\sigma$, $3$$\sigma$ and $5$$\sigma$ prospective sensitivities of the $1.3~(3.5)$ TeV LHeC (FCC-eh) to the ALP-lepton couplings $Tr(g_{a\ell\ell})/f_a$. Our numerical results show that the $3.5$ TeV FCC-eh has greater potential to explore the ALP-lepton couplings than the $1.3$ TeV LHeC.
The expected limits on the ALP-lepton couplings for the ALPs with masses in the range of $1$ $\sim$ $5$ GeV via the process $e^- p \to e^- j a~(a \to \mu^+ \mu^-)$ are stronger than those from the process $e^{-}p\rightarrow e^-ja~(a \to \tau^+ \tau^-, \tau^+ \tau^- \to \ell^+ \ell^{-} \slashed E, \ell= e,\mu)$ for the ALPs with masses in the range of $5$ $\sim$ $500$ GeV at the $1.3~(3.5)$ TeV LHeC~(FCC-eh) for the cases of $c_R = 0$ and $c_L = 0$. Furthermore, the $1.3~(3.5)$ TeV LHeC (FCC-eh) is capable of achieving better sensitivities to the ALP-lepton couplings via all signal processes considered in this paper for $c_L = 0$ than those for $c_R = 0$.

Comparing our numerical results with other existing bounds and prospective sensitivities, we find that the $1.3~(3.5)$ TeV LHeC (FCC-eh) can cover a part of the ALP parameter space that are not excluded by current and future experiments. Specifically, the prospective sensitivities of the $1.3~(3.5)$ TeV LHeC (FCC-eh) to the ALP-lepton coupling $Tr(g_{a\ell\ell})/f_a$ are in the ranges of $0.110~(0.081)$ $\sim$ $0.440~(0.160)$ GeV$^{-1}$ with $120~(110)$ GeV $\leq$ $m_a$ $\leq$ $500$ GeV for $c_{R}=0$ and $0.092~(0.042)$ $\sim$ $0.310~(0.070)$ GeV$^{-1}$ with $120~(45)$ GeV $\leq$ $m_a$ $\leq$ $500$ GeV for $c_{L}=0$ via the process $e^{-}p\rightarrow e^-ja~(a \to \tau^+ \tau^-, \tau^+ \tau^- \to \ell^+ \ell^{-} \slashed E, \ell= e,\mu)$. In addition, for $c_{L}=0$ and $2.5~(1)$ GeV $\leq$ $m_a$ $\leq$ $4~(5)$ GeV, the values of $Tr(g_{a\ell\ell})/f_a$ in the range of $0.011~(0.001)$ $\sim$ $0.112~(0.040)$ GeV$^{-1}$ obtained via the process $e^{-}p\rightarrow e^-ja~(a \to \mu^+ \mu^-)$ are also not excluded by current experiments.

In conclusion, our estimates of the prospective sensitivities of the LHeC~(FCC-eh) are not only stronger than some existing bounds of the LHC and LEP but also complementary to the expected bounds of the CEPC and FCC-ee. Thus, we can say the LHeC~(FCC-eh) has a better potential to detect the ALP-lepton couplings or give more severe projected sensitivities. Hopefully this work will provide valuable insights for future collider experiments aiming to explore the ALP-lepton couplings.

\textbf{Note added.} When finishing this work, we find that the constraints from supernova observations have been updated in Refs.~\cite{Fiorillo:2025yzf,Fiorillo:2025sln}.
\section*{ACKNOWLEDGMENT}

This work was partially supported by the National Natural Science Foundation of China under Grants No. 11875157 and No. 12147214.


\end{document}